\documentclass[times,twocolumn]{aastex62}
\usepackage{amsmath}
\usepackage{amssymb}

\def\vector#1{\mbox{\boldmath $#1$}}
\newcommand{\dpar}[2]{\frac{\partial#1}{\partial#2}}
\newcommand{\bfrac}[3]{\left(\frac{#1}{#2}\right)^{#3}}
\newcommand{\mstar}{M_{\ast}}
\newcommand{\vgas}{\vector{V}}
\newcommand{\vrgas}{V_{\rm R}}
\newcommand{\vtgas}{V_{\phi}}
\newcommand{\vdust}{\vector{v}}
\newcommand{\vrdust}{v_{\rm R}}
\newcommand{\vtdust}{v_{\rm \phi}}
\newcommand{\sigmagas}{\Sigma_{\rm g}}
\newcommand{\sigmadust}{\Sigma_{\rm d}}
\newcommand{\tstop}{t_{\rm stop}}
\newcommand{\dsize}{s_{\rm d}}
\newcommand{\dsizefrag}{s_{\rm frag}}
\newcommand{\dsizedrift}{s_{\rm drift}}
\newcommand{\dsizeo}{s_{\rm d,0}}
\newcommand{\rhop}{\rho_p}
\newcommand{\Omegak}{\Omega_{\rm K}}
\newcommand{\sonic}{c_s}
\newcommand{\st}{St}

\newcommand{\hg}{h_{\rm g}}
\newcommand{\rp}{R_{\rm p}}
\newcommand{\mpl}{M_{\rm p}}
\newcommand{\grav}{G}

\newcommand{\au}{\ \rm{AU}}
\newcommand{\wring}{\Delta_{\rm ring}}
\newcommand{\rmean}{R_{\rm m}}
\newcommand{\hgp}{h_{\rm g,p}}
\newcommand{\dgratioth}{(\Sigma_{\rm d}/\Sigma_{\rm g})_{\rm th}}
\newcommand{\ffrag}{f_{\rm frag}}
\newcommand{\fdrift}{f_{\rm drift}}
\newcommand{\ufrag}{u_{\rm frag}}
\newcommand{\sigmaunit}{\mbox{ g/cm}^2}

\newcommand{\RED}[1]{#1}
\received{\today}
\revised{xxxx}
\accepted{xxxx}
\submitjournal{ApJ}

\shorttitle{Impacts of dust feedback on a dust ring induced by a planet}
\shortauthors{Kanagawa et al.}
\begin{document}

\title{Impacts of dust feedback on a dust ring induced by a planet in a protoplanetary disk}

\correspondingauthor{Kazuhiro D. Kanagawa}
\email{kazuhiro.kanagawa@utap.phys.s.u-tokyo.ac.jp}

\author[0000-0001-7235-2417]{Kazuhiro D. Kanagawa}
\affiliation{Research Center for the Early Universe, Graduate School of Science, University of Tokyo, Hongo, Bunkyo-ku, Tokyo 113-0033, Japan}
\affiliation{Institute of Physics and CASA$^{\ast}$, Faculty of Mathematics and Physics, University of Szczecin, Wielkopolska 15, PL-70-451 Szczecin, Poland}

\author{Takayuki Muto}
\affiliation{Division of Liberal Arts, Kogakuin University,1-24-2 Nishi-Shinjuku, Shinjuku-ku, Tokyo 163-8677, Japan}

\author{Satoshi Okuzumi}
\affiliation{Department of Earth and Planetary Sciences, Tokyo Institute of Technology, 2-12-1 Ookayama, Meguro, Tokyo, 152-8551, Japan}

\author{Takayuki Tanigawa}
\affiliation{National Institute of Technology, Ichinoseki College, Ichinoseki-shi, Iwate 021-8511, Japan}

\author{Tetsuo Taki}
\affiliation{School of Arts \& Sciences, University of Tokyo, 3-8-1, Komaba, Meguro, 153-8902 Tokyo, Japan}
\affiliation{Center for Computational Astrophysics, National Astronomical Observatory of Japan, Osawa, Mitaka, Tokyo 181-8588, Japan}

\author{Yuhito Shibaike}
\affiliation{Department of Earth and Planetary Sciences, Tokyo Institute of Technology, 2-12-1 Ookayama, Meguro, Tokyo, 152-8551, Japan}


\begin{abstract}
When a planet forms a deep gap in a protoplanetary disk, dust grains cannot pass through the gap.
As a consequence, the density of the dust grains can increase up to the same level of the density of the gas at the outer edge.
The feedback on the gas from the drifting dust grains is not negligible, in such a dusty region.
We carried out two-dimensional two-fluid (gas and dust) hydrodynamic simulations.
We found that when the radial flow of the dust grains across the gap is halted, a broad ring of the dust grains can be formed because of the dust feedback and the diffusion of the dust grains.
The minimum mass of the planet to form the broad dust ring is consistent with the pebble-isolation mass, in the parameter range of our simulations.
The broad ring of the dust grains is good environment for the formation of the protoplanetary solid core.
If the ring is formed in the disk around the sun-like star at $\sim 2\au$, a massive solid core ($\sim 50M_{\oplus}$) can be formed within the ring, which may be connected to the formation of Hot Jupiters holding a massive solid core such as HD~149026b.
In the disk of the dwarf star, a number of Earth-sized planets can be formed within the dust ring around $\sim 0.5\au$, which potentially explain the planet system made of multiple Earth-sized planets around the dwarf star such as TRAPPIST-1.
\end{abstract}



\section{Introduction} \label{sec:intro}
In a protoplanetary disk, a giant planet such as Jupiter can form a density gap due to strong disk--planet interaction along with its orbit \citep[e.g.,][]{Lin_Papaloizou1979,Goldreich_Tremaine1980,Lin_Papaloizou1986a}.
Recent observations have discovered bright rings and dark gaps of the dust grains \citep[e.g.,][]{Fukagawa2013,vanderMarel2013,Perez2014,Muto2015, ALMA_HLTau2015,Akiyama2015,Momose2015,Nomura_etal2016,Tsukagoshi2016,Akiyama2016,Isella2016,Kataoka2016,van_der_Plas_etal2017,Fedele_etal2017,Dong2018}.
These structures can be associated with the planets embedded within the protoplanetary disk \citep[e.g.,][]{Dong_Zhu_Whitney2015,Kanagawa2015b,Dipierro_Price_Laibe_Hirsh_Cerioli_Lodato2015,Picogna_Kley2015,Jin_Li_Isella_Li_Ji2016,Rosotti_Juhasz_Booth_Clarke2016,Dipierro_etal2018}.

A planet is formed by accumulation of dust grains in the protoplanetary disk.
Because of the importance of the dust grains, the evolution of the dust grains in the disk have been investigated \citep[e.g.,][]{Nakagawa_Sekiya_Hayashi1986,Youdin_Shu2002,Tanaka_Himeno_Ida2005,Birnstiel_Klahr_Ercolano2012,Okuzumi_Tanaka_Kobayashi_Wada2012,Drazkowska_Alibert_Moore2016,Ida_Guillot2016}.
When the planet forms the density gap, a relatively large sized dust grains (which is so-called by pebble) are trapped by the pressure bump which is formed at the outer edge of the gap.
As a result, the ring structure in which the dust grains are highly concentrated can be formed at the outer edge \citep[e.g.,][]{paardekooper2004,Muto_Inutsuka2009b,Zhu2012,Dong_Zhu_Whitney2015,Pinilla_Ovelar_Ataiee_Benisty_Birnstiel_Dishoeck_Min2015,Pinilla_Klarmann_Birnstiel_Benisty_Dominik_Dullemond2016}.
The dust grains move inward due to friction with the surrounding disk gas, and at the same time, the disk gas feels feedback from the dust grains.
Although the feedback is negligible if the dust grains are small and do not piled-up, we need to consider effects of the feedback if the dust grains are relatively large or highly concentrated.
In this case, the dust feedback can significantly influence the structure of the disk gas \citep{Fu2014,Gonzalez2015,Taki_Fujimoto_Ida2016,Gonzalez2017,Dipierro_Laibe2017,Kanagawa_Ueda_Muto_Okuzumi2017,Weber2018,Dipierro_Laibe_Alexander_Hutchison2018}.
%Recently, by performing two-fluid two-dimensional hydrodynamic simulations, \cite{Weber2018} have found that the dust feedback does not significantly affect the structure within the gap.
%However, the dust feedback significantly affects the ring structure formed in the outer disk of the planet.
Recently, by performing two-fluid two-dimensional hydrodynamic simulations, \cite{Weber2018} have found that although the dust feedback does not significantly affect the total amount of dust transported through the planetary gap, it substantially changes the location at which the dust accumulates outside of the planet's orbit.

Within the ring structure of the dust grains in which the dust grains are highly concentrated, the planetesimals and the solid protoplanetary core can be effectively produced.
The formation of the solid core in the ring structure with the high concentration of the dust grains has been investigated by several previous studies \citep[e.g.,][]{Kobayashi_Ormel_Ida2012,Pinilla_Ovelar_Ataiee_Benisty_Birnstiel_Dishoeck_Min2015}.
However, the effects of the dust feedback had not been considered in the previous studies.
The effects of the dust feedback on the ring structure need to be investigated, which might be connected to the formation of some strange planets which are difficult to be understood by the current theoretical framework.
For instance, compact Hot Jupiters which have a small radius as compared with its mass (that is, with a high mean density), e.g., HD~149026b, have been observed \citep{Sato2005,Hebrard2013}.
Such a Hot Jupiter is thought to have a massive solid core as $\sim 50 M_{\oplus}$ \citep{Fortney_Saumon_Marley_Lodders_Freedman2006,Ikoma_Guillot_Genda_Tanigawa_Ida2006}.
Only in the environment with the very high density of the dust grains, these Hot Jupiters can be formed \citep{Ikoma_Guillot_Genda_Tanigawa_Ida2006}.
Moreover, recent observations have revealed several Earth-sized planets orbiting around the dwarf star named TRAPPIST-1 \citep{Gillon2016,Gillon2017}.
Since the mass of the solid materials would be small in the protoplanetary disk around the dwarf star, it is difficult to understand the formation of a multitude of planets in such a disk.
Recently \cite{Ormel_Liu_Schoonenberg2017} have shown that if the disk of the dwarf star extends to the same extent as that of the sun-like star ($\sim 100\au$), several Earth-sized planet can be formed around the water-snow line.
Alternatively, \cite{Haworth_Facchini_Clarke_Mohanty2018} have shown that it is possible if the mass of the disk is massive.
If a wide ring of the dust grains is formed in the outside of the gap, however, a multitude of Earth-sized planets may be able to form within this ring, even when the disk is compact and less massive.

In this paper, we investigate the effects of the dust feedback on the ring structure of the dust grains which is formed at the outer edge of the planet-induced gap, by using two-fluid (gas and dust) two-dimensional hydrodynamic simulations.
In Section~\ref{sec:basic_eq}, we describe the basic equations.
We show the result of the hydrodynamic simulations in Section~\ref{sec:results}.
In Section~\ref{sec:discussion}, we discuss the effects of the planetesimal formation on the ring structure and implications on the planet formation within the dust ring and observations.
We also discuss the validity of the our simulations in this section.
Our summary is included in Section~\ref{sec:summary}.

\section{Basic equations and numerical method} \label{sec:basic_eq}
\subsection{Basic equations for dust grains and disk gas} \label{eq:beq}
We simulate the structures of disk gas and dust grains in the protoplanetary disk with the planet.
In this paper, we assume a geometrically thin and non-self gravitating disk.
We do not consider a detail of the vertical structure of the disk, and we deal with the vertical averaged values of physical quantities, such as the surface density defined by $\Sigma = \int^{\infty}_{-\infty} \rho dz$, where $\rho$ is a density of gas or dust grains.
We use two-dimensional ($R,\phi$) coordinate, and its origin is set on the position of the central star.
The gas and dust velocities are expressed by $\vgas=(\vrgas,\vtgas)$ and $\vdust=(\vrdust,\vtdust)$, respectively, and the surface densities of the gas and the dust grains are  written by $\sigmagas$ and $\sigmadust$, respectively.
In the following, the subscripts, $g$ and $d$, indicate values for gas and dust grains, respectively.
We treat dust grains as a pressureless fluid.

The equations of motions for dust grains in radial and azimuthal directions are given by
\begin{align}
&\dpar{\vrdust}{t}+\left(\vdust \cdot \nabla \right) \vrdust -\frac{\vtdust^2}{R} = -\dpar{\Psi}{R} - \frac{\vrdust - \vrgas}{\tstop} \label{eq:dust_eom_radial},\\
&\dpar{\vtdust}{t}+\left(\vdust \cdot \nabla \right) \vtdust + \frac{\vrdust \vtdust}{R}  = -\frac{1}{R}\dpar{\Psi}{\phi} - \frac{\vtdust - \vtgas}{\tstop} \label{eq:dust_eom_azimuthal}.
\end{align}
The last terms in RHS of equations~(\ref{eq:dust_eom_radial}) and (\ref{eq:dust_eom_azimuthal}) represent drag forces between disk gas and dust grains.
The stopping time of dust grains, $\tstop$, depends on size of the dust grains.
For simplicity, we consider only dust grains in the Epstein regime, which is reasonable to the range of the dust grains considered in this paper.
In this case, the stopping time is written by \citep{Takeuchi_Lin2005}
\begin{align}
&\tstop = \frac{\pi \dsize \rhop}{2\sigmagas \Omegak},
\end{align}
where $\dsize$ and $\rhop$ are a size and an internal density of the dust grains, and $\Omegak = \sqrt{G\mstar/R^3}$ is the Keplerian angular velocity, where $\grav$ is the gravitational constant, $\mstar$ is the mass of the central star, respectively.
For convenience, we define the Stokes number of the dust grains as
\begin{align}
&\st=\tstop \Omegak \label{eq:st}.
\end{align}
The gravitational potential $\Psi$ is given by
\begin{align}
	\Psi= & - \frac{\grav \mpl}{\left[ R^2+2R\rp \cos\left( \phi-\phi_p \right) + \rp^2 +\epsilon^2  \right]^{1/2}} \nonumber \\
	&  \qquad  \qquad \qquad \quad -\frac{\grav \mstar}{R} + \frac{\grav \mpl}{\rp^2} R\cos\left( \phi-\phi_p \right),\label{eq:gravpot}
\end{align}
where $\mpl$ is the mass of the planet, which is located at ($\rp$,$\phi_p$), where $\rp$ and $\phi_p$ are the orbital radius of the planet and the azimuthal angle of the planet, respectively.
The softening parameter is denoted by $\epsilon$.
For disk gas, the equations of motion in the radial and azimuthal direction are
\begin{align}
&\dpar{\vrgas}{t}+\left(\vgas \cdot \nabla \right)\vrgas -\frac{\vtgas^2}{R} = -\frac{\sonic^2}{\sigmagas} \dpar{\sigmagas}{R} - \dpar{\Psi}{R} \nonumber \\
& \qquad \qquad \qquad \qquad \qquad \qquad \qquad + \frac{f_R}{\sigmagas} - \frac{\sigmadust}{\sigmagas} \frac{\vrgas - \vrdust}{\tstop} \label{eq:gas_eom_radial} ,\\
&\dpar{\vtgas}{t} + \left(\vgas \cdot \nabla \right) \vtgas + \frac{\vrgas \vtgas}{R} = -\frac{\sonic^2}{\sigmagas} \dpar{\sigmagas}{\phi} - \frac{1}{R}\dpar{\Psi}{R} \nonumber \\
& \qquad \qquad \qquad \qquad \qquad \qquad \qquad + \frac{f_{\phi}}{\sigmagas} - \frac{\sigmadust}{\sigmagas} \frac{\vtgas - \vtdust}{\tstop} \label{eq:gas_eom_azimuthal},
\end{align}
where we adopted a simple isothermal equation of state, in which a vertical averaged pressure is given by $\sonic^2 \sigmagas$, where $\sonic$ is an isothermal sound speed.
The viscous forces in radial and azimuthal directions are represented by $f_R$ and $f_{\phi}$, respectively \citep[c.f.,][]{Nelson_Papaloizou_Masset_Kley2000}.
We adopt $\alpha$--prescription \citep{Shakura_Sunyaev1973}, and hence the kinetic viscosity $\nu$ is expressed by $\alpha \hg^2 \Omegak$, where $\hg$ is a disk scale height of the gas.
In this paper, we assume a constant $\alpha$ throughout the disk.

The continuity equation for the gas in the two-dimensional disk is written by
\begin{align}
&\dpar{\sigmagas}{t}+\nabla \cdot \left(\sigmagas \vgas \right) = 0 \label{eq:continuity_gas}.
\end{align}
The turbulence in the gas disk drives random motion of dust grains, which induces diffusion of the dust grains \citep[e.g.,][]{Cuzzi_Dobrovolskis_Champney1993,Youdin_Lithwick2007}.
Hence, considering this turbulence diffusion of the dust grains, we obtain the vertically averaged continuity equation of dust grains as
\begin{align}
&\dpar{\sigmadust}{t}+\nabla \cdot \vector{F}_{\rm M,d} = 0 \label{eq:continuity_dust},
\end{align}
where a mass flux of the dust grains is
\begin{align}
\vector{F}_{\rm M,d} &= \sigmadust \vdust + \vector{j} 
\label{eq:dust_mass_flux}
\end{align}
where $\vector{j}$ represents a diffusive mass flux.
Assuming diffusion caused by the spatial gradient (Gradient diffusion hypothesis), we can describe the diffusive mass flux as
\citep[e.g.,][]{Cuzzi_Dobrovolskis_Champney1993,Takeuchi_Lin2002,Zhu2012}
\begin{align}
&\vector{j}= -D \sigmagas \nabla \left(\frac{\sigmadust}{\sigmagas}\right),
\label{eq:j}
\end{align}
where $D$ is the turbulent diffusivity of the dust grains.
The gradient diffusion hypothesis is appropriate for the dust grains with small $\st$ \citep[e.g.,][]{Fromang_Papaloizou2006,Okuzumi_Hirose2011}.
This hypothesis may be inappropriate if the dust grains have the size of $\st \sim 1$ in a highly dust-rich region as $\sigmadust \sim \sigmagas$.
In this case, we should directly calculate an interaction between the dust grains and turbulence in gas.
However, the motion of the dust grains in such a situation is poorly understood yet.
Hence, in this paper, we adopt the model of the diffusion of the dust grains given by Equations~(\ref{eq:dust_mass_flux}) and (\ref{eq:j}), regardless of the values of $\st$ and $\sigmadust/\sigmagas$.
In this case, using the model of \cite{Youdin_Lithwick2007}, we obtain the turbulent diffusivity of the dust grains as
\begin{align}
D&=\nu \frac{1+4\st^2}{\left(1+\st^2 \right)^2}.
\label{eq:turbulent_diffusivity}
\end{align}

\subsection{Size evolution of the dust grains} \label{subsec:dust_size_evo}
In this paper, we consider two cases: one is the case in which the size of the dust grains is constant during the simulations, and another is the case in which the growth of the dust grains is considered.
Here we describe the model of the dust growth which we adopted in our simulations.
The size of the dust grains varies by e.g., the coagulation and collisional fragmentation \citep[e.g.,][]{Birnstiel_Dullemond_Brauer2010,Okuzumi_Tanaka_Kobayashi_Wada2012}.
Although the dust grains has a size distribution, \cite{Birnstiel_Klahr_Ercolano2012} have provided the simple model of the dust size evolution with a single representative size of the dust grains.
\RED{
Their model does not consider the situation with a planet and neglects some physical processes such as the feedback from the gas, detail vertical dynamics and turbulent statistics of growing grains (see also Section~\ref{subsection:validty_model}).
However, their model is useful for us as the first step, to include the size evolution of the dust grains by considering the single representative size of the dust grain, instead of directly calculating the coagulation and fragmentation of the dust grains.
}
Recently \cite{Tamfal_Drazkowska_Mayer_Surville2018} have considered the dust growth in the two-dimensional hydrodynamic simulations by using the similar way.
In the following, we briefly summarize the size evolution model of the dust grains provided by \cite{Birnstiel_Klahr_Ercolano2012}.

The size of the dust grains can be characterized by the maximum size of the dust grains.
When it is determined by the fragmentation due to the turbulence, the stokes number of the maximum sized-grains is obtained by $\simeq \alpha^{-1}(\ufrag/\sonic)^2 $, where $\ufrag$ is the fragmentation threshold velocity.
That is, in the Epstein regime,  the maximum size of the dust grains can be given by 
\begin{align}
\dsizefrag &= \ffrag \frac{2}{3\pi} \frac{\sigmagas}{\rhop \alpha} \left(\frac{\ufrag}{\sonic}\right)^2,
\label{eq:size_frag}
\end{align}
where $\ffrag$ is a correction factor which is given by $0.37$.

Radial drift can limit the maximum size of the dust grains if the timescale of the dust radial drift is comparable with, or less than, the timescale of the dust growth.
The timescale of the dust growth can be estimated by
\begin{align}
\tau_{\rm growth} &= \frac{1}{\Omegak} \left(\frac{\sigmagas}{\sigmadust} \right).
\label{eq:tau_growth}
\end{align}
\RED{
For dust grains with $\st<1$, the timescale of the dust radial drift can be estimated by
\begin{align}
\tau_{\rm drift} &= \frac{R}{\left|\vrdust \right|} = \frac{1}{\st} \left(\frac{\hg}{R}\right)^{-2} \left|\frac{\partial \ln \left(\sonic^2 \sigmagas \right)}{\partial \ln R} \right|^{-1} \frac{1}{\Omegak}.
\label{eq:tau_drift}
\end{align}
}
When the radial drift dominates the size of the dust grains, the maximum size of the dust grains can be given by $\tau_{\rm growth} \simeq \tau_{\rm drift}$.
That is, using the Epstein law, we obtain
\RED{
\begin{align}
\dsizedrift &= \fdrift \frac{2}{\pi} \frac{\sigmadust}{\rhop} \left(\frac{\hg}{R} \right)^{-2} \left|\frac{\partial \ln \left(\sonic^2 \sigmagas \right)}{\partial \ln R} \right|^{-1},
\label{eq:size_drift}
\end{align}
}
where the correction factor $\fdrift$ is $0.55$.

The representative size of the dust grains takes the smaller of $\dsizefrag$ and $\dsizedrift$.
In the model developed by \cite{Birnstiel_Klahr_Ercolano2012}, the size growth of the dust grains from the size of monomers is considered.
However, since we assume the situations in which the planet is already formed, it could be reasonable to consider that the size of the dust grains reaches either size given by Equations~(\ref{eq:size_frag}) or (\ref{eq:size_drift}).
Hence, we describe the representative size of the dust grains as
\begin{align}
\dsize(R,t) &= \min \left[ \dsizefrag, \dsizedrift \right],
\label{eq:size_representative}
\end{align}

Note that in the model described above, we assume the size distribution of the dust grains is in coagulation--fragmentation equilibrium when the representative size of the dust grains is determined by the fragmentation.
In the unperturbed disk, this assumption would be valid.
However, when the density of the dust grains is significantly small and then the dust growth time is very long (e.g., in the edge of the gap), this assumption may not be valid.
The above situation could be realized in the inner disk of the planet when the gap is very deep.
According to Equation~(\ref{eq:size_representative}), the represent size of the dust grains is very small due to $\dsizedrift$ with a very small value of $\sigmadust$.
Hence, since the stopping time of the dust grains becomes short, we must choose a smaller time step.
To avoid the above condition, we keep the representative size of the dust grains to be $\st=0.1$ by use the function of $\st+0.1\exp(\sigmadust/[10^{-4}\sigmagas])$, if the dust-to-gas mass ratio is smaller than $10^{-4}$.
Because we focus on the structure of the outer disk in this paper, this prescription does not affect our results.

It should also be noted that we consider only the collision velocity caused by the gas turbulence, which is reasonable if the value of $\alpha$ is relatively large.
However, if the situation with a very small value of $\alpha$ is considered, the collision velocity of the dust grains is dominated by the difference of the radial drift velocities, as noted by \cite{Birnstiel_Klahr_Ercolano2012}.
In this case, we cannot adopt Equation~(\ref{eq:size_frag}) for $\dsizefrag$.
In our parameter range of $\alpha \geq 10^{-3}$, the collision velocity related to the radial drift is smaller than or comparable to that related to the turbulence.
Hence, we neglect this effect in this paper \footnote{We estimated the collision velocity related to the radial drift at each time step and in each spatial mesh. If it is larger than that related to the turbulence, we determined $\dsizefrag$ by the collision velocity related to the radial velocity, instead of that related to the turbulence. However, in most cases, $\dsizefrag$ was given by Equation~(\ref{eq:size_frag}) in our parameter range.
In the case of smaller $\alpha$, the fragmentation induced by the radial drift could be significant.}.
To include the size evolution of the dust grains precisely, more sophisticated model \citep[e.g.,][]{Sato_Okuzumi_Ida2016} is required.

\subsection{Numerical method and setup} \label{subsec:numerical_method}
We carried out two-fluid hydrodynamic simulations by a code based on {\sc \tt FARGO} \citep{Masset2000}, which is an Eulerian polar grid code with a staggered mesh.
The FARGO algorithm, which removes the azimuthally averaged velocity for the Courant time step, enables us long--term simulations.
Expending {\sc \tt FARGO} to include the dust component, we numerically solved the equations of motions and the continuity equations for gas and dust grains \citep{Kanagawa_Ueda_Muto_Okuzumi2017}.

The computational domain runs from $R/\rp=0.4$ to $R/\rp=4.0$ with $512$ radial zones (equally spaced in logarithmic space) and $1024$ azimuthal zones (equally spaced).
The subscript $\rm{p}$ indicates the value at $R=\rp$ in the following.
The initial condition of the surface density of the disk gas is set as $\sigmagas=\Sigma_0 (R/\rp)^{-1}$ and $\Sigma_0=6\times10^{-4}$, which corresponds $53 \sigmaunit$ when $\mstar=1M_{\odot}$ and $\rp=10\au$.
The disk aspect ratio is assumed to be $\hg/R = (\hgp/\rp) (R/\rp)^{1/4}$ with $\hgp/\rp=0.05$.
The initial angular velocity of the disk gas is given by $\Omegak\sqrt{1-\eta}$, where 
\RED{
\begin{align}
\eta = -\frac{1}{2}\bfrac{\hg}{R}{2} \frac{\partial \ln \left(\sonic^2 \sigmagas \right)}{\partial \ln R}.
\label{eq:eta}
\end{align}
}
For the dust component, the initial distribution of the dust grains is set for the dust-to-gas mass ratio to be $0.01$.
The initial angular velocity of the dust grains is given by $\Omegak$ and the initial radial velocity is set to be zero.
For simplicity, we neglect mass growth of the planet.
We also neglect time-variation of the planetary orbit, and the planets' orbit is fixed to be $R=\rp$.
For convenience, we define $t_0$ as $2\pi/\Omegak{}_{\rm,p}$.
We adopt the softening parameter $\epsilon$ as $0.6\hgp$.
We use the orbital radius of the planet $\rp$ as the unit of the radius, and the mass of the central star $\mstar$ is used as the unit of the mass.
Hence, the surface density is normalized by the value of $\mstar/\rp^2$.

At the inner and outer boundaries, the gas and dust velocities are set to be these in steady state during the simulations.
The surface densities of the gas and dust are also set to that the mass flux is constant.
To avoid an artificial wave reflection, the damping is used in a so-called wave-killing zone near the boundary layers as in \cite{Val-Borro_etal2006}.
In this region, we force the physical quantities to be azimuthally symmetric.
For detail, see \cite{Kanagawa_Ueda_Muto_Okuzumi2017}. 
The wave-killing zones are located from $R_{\rm out}-0.2\rp$ to $R_{\rm out}$ for the outer boundary and from $R_{\rm in}$ to $R_{\rm in}+0.1\rp$ for the inner boundary, where $R_{\rm out}$ and $R_{\rm in}$ are the radius of the outer and inner boundaries, respectively.

We carried out two kinds of simulations: One considers the size evolution of the dust grains by using the model described in Section~\ref{subsec:dust_size_evo}, another does not consider the size evolution of the dust grains.
When the size evolution of the dust grains is not considered, the size of the dust grains is constant throughout the disk during the simulations.
For convenience, we define a characteristic size of the dust grain as $\dsizeo=2\Sigma_0/(\pi \rhop)$, and that is,
\begin{align}
\dsizeo &= 34.0 \bfrac{\sigmagas{}_{\rm ,un}}{53 \sigmaunit}{} \bfrac{\rhop}{1\mbox{ g/cm}^3}{-1} \mbox{ cm},
\end{align}
where $\sigmagas{}_{\rm ,un}$ is the gas surface density of the unperturbed disk.
The size of $\dsizeo$ corresponds to the size of the dust grain when $\st = 1$ at $R=\rp$ in the unperturbed disk.
We usually adopt $\dsize = 0.1 \dsizeo$ when the size evolution of the dust grains is not considered.

When the size evolution of the dust grains is considered, the size of the dust grains varies as described by the model described in Section~\ref{subsec:dust_size_evo}.
In this case, the maximum size of the dust grains is associated to the threshold velocity of the fragmentation $\ufrag$, if it is determined by the fragmentation.
The threshold velocity depends on the composition of the dust grains.
For the icy grains, the threshold velocity of the fragmentation could be as high as $50 \mbox{m/s}$ \citep[e.g.,][]{Wada_Tanaka_Suyama_Kimura_Yamamoto2009}, when the size of the monomer is $0.1\mu \mbox{m}$.
For the rocky grains, the threshold velocity of the fragmentation is about $\sim 1 \mbox{m/s}$, which is much smaller than that for the icy grains.
However, since the threshold velocity depends on the size of the monomers \citep[e.g.,][]{Dominik_Tielens1997}, the threshold velocity could be as the same level as that for the icy grains, if the nanograins is assumed \citep{Arakawa_Nakamoto2016}.
In this paper, therefore, we consider $\ufrag = 30 \mbox{m/s}$ as a fiducial value.

\section{Results of hydrodynamic simulations} \label{sec:results}
\subsection{Ring structures induced by Jupiter-sized planets} \label{subsec:ring_jupiter}
\subsubsection{Density distributions}
\begin{figure*}
	\begin{center}
		\resizebox{0.98\textwidth}{!}{\includegraphics{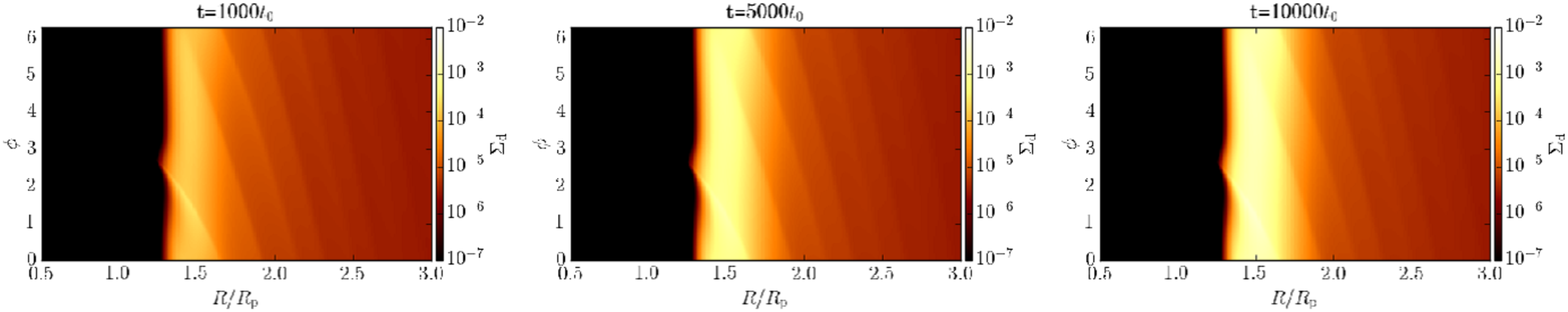}}
		\resizebox{0.98\textwidth}{!}{\includegraphics{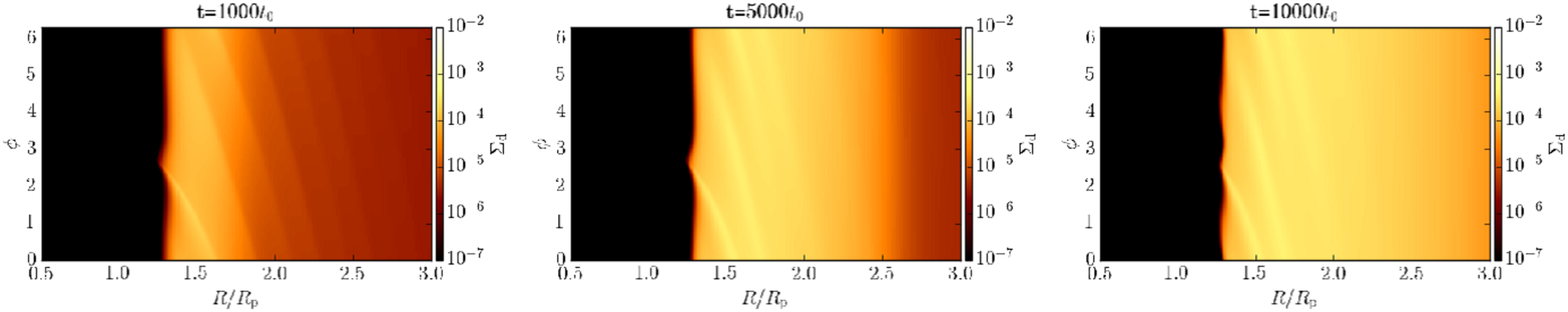}}
		\caption{The two-dimensional distributions of the dust surface density with the dust grains of $\dsize = 0.1 \dsizeo$ at $t=1000t_0$ (left), $5000t_0$ (middle), and $10000t_0$ (right), when $\mpl/\mstar = 10^{-3}$, $\hgp/\rp=0.05$, and $\alpha=4\times 10^{-3}$.
		In the upper panels, the dust feedback is ignored, while the feedback is considered in the lower panels.
		The dust growth is not considered in both the cases.
		\label{fig:rings_q1e-3_a4e-3_h0.05_const_dustsize}
		}
	\end{center}
\end{figure*}
First we show the two-dimensional distributions of the surface density of the dust grains, in the case with the constant size of the dust grains ($\dsize = 0.1\dsizeo$), in Figure~\ref{fig:rings_q1e-3_a4e-3_h0.05_const_dustsize}.
In this subsection, we show the results of the simulations with the Jupiter-sized planet ($\mpl/\mstar=10^{-3}$), and we adopt $\hgp/\rp = 0.05$ and $\alpha=4\times 10^{-3}$ as fiducial values.
In the upper panel of the figure, the dust feedback is not considered.
In the lower panel of the figure, we consider the dust feedback.
In both the cases with and without the dust feedback, the radial flow of the dust grains across the gap is halted, because the gap of the gas is deep (the ratio of the minimum surface density of the gap to the unperturbed surface density is about $\sim 0.01$).
Hence, the dust grains are completely swept out from the region within the gap and the inside of the planet orbit.
As a consequence, a ring structure of the dust grains is formed in the outer disk of the planet orbit.
In the earlier stage as in the case of $t=1000t_0$, the widths of the dust ring are almost the same in both the cases in which the dust feedback is considered and it is ignored.
In the case in which the dust feedback is considered, however, the dust ring is gradually wider with time, whereas the width of the dust ring in the case in which the feedback is ignored is much narrower and no longer changes after $1000t_0$.

We should note that when the dust feedback is considered, the distributions of the gas surface density is slightly modified from that in the case without the dust feedback as in the distribution of the dust grains.
The dust feedback reduces the gas surface density by a factor at the outer edge of the gap, as shown in Appendix~\ref{sec:gas_disk_structures}.
\RED{
We also examine the resolution dependence of the gas and dust structures in this case (see Appendix~\ref{sec:resolution_convergence}), and find that with our fiducial resolution, the structures of the gas and the dust grains are well converged.
}

\begin{figure*}
	\begin{center}
		%\resizebox{0.98\textwidth}{!}{\includegraphics{gap_with_feedback_a4e-3_h0.05_q1e-3_St0.01Init_DustGrowth_512x1024v2_dust0_2ddists.eps}}
		\resizebox{0.98\textwidth}{!}{\includegraphics{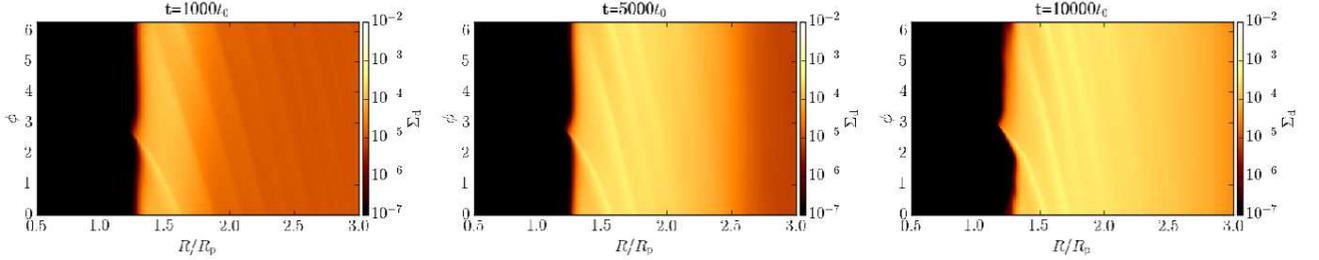}}
		\caption{
		The same as Figure~\ref{fig:rings_q1e-3_a4e-3_h0.05_const_dustsize}, but in this case, the size evolution of the dust grains is considered.
		\label{fig:rings_q1e-3_a4e-3_h0.05_wDustgrowth}
		}
	\end{center}
\end{figure*}
\begin{figure}
	\begin{center}
		\resizebox{0.49\textwidth}{!}{\includegraphics{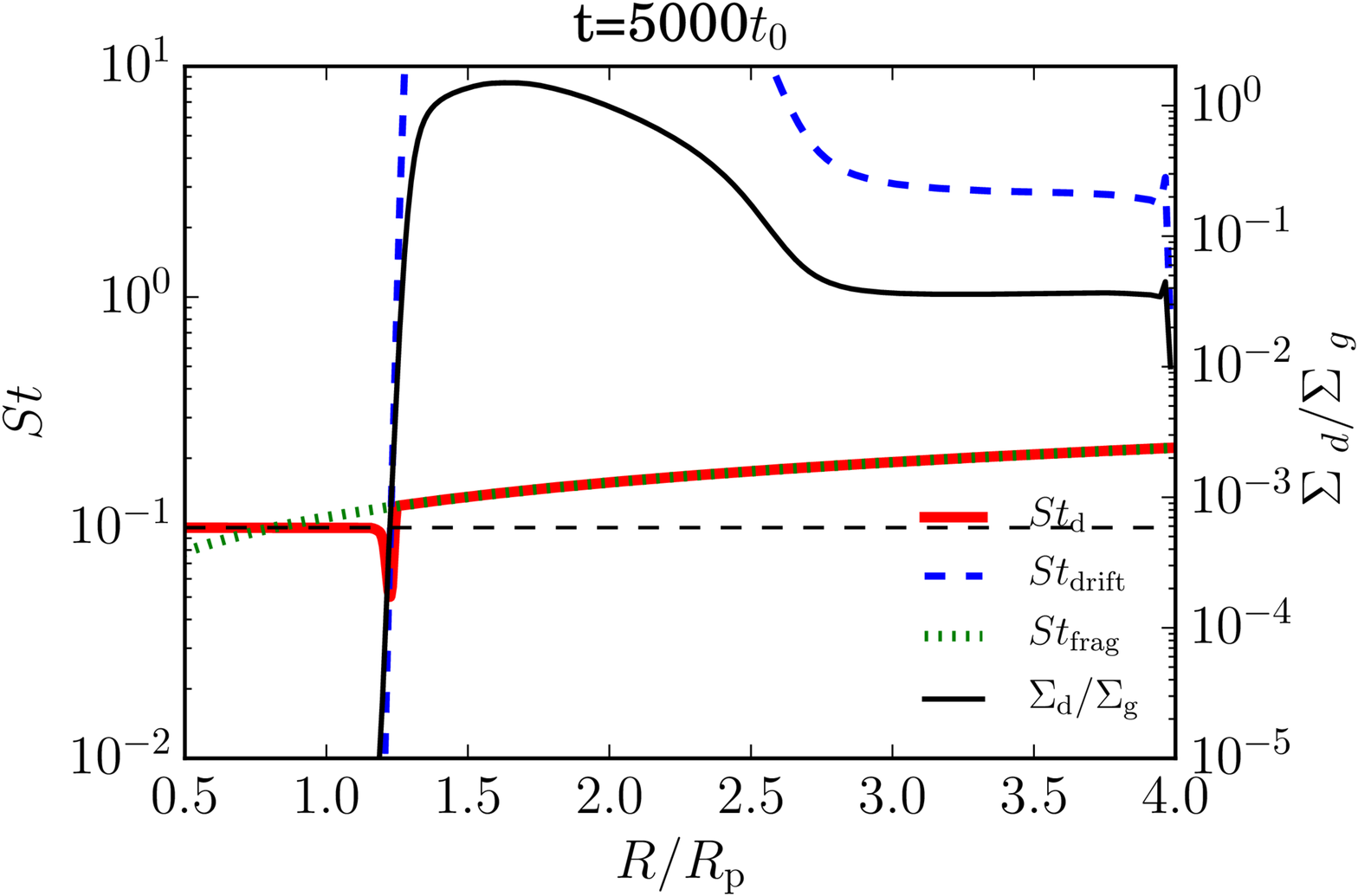}}
		\caption{
		The distribution of the stokes number for the representative size in the case of Figure~\ref{fig:rings_q1e-3_a4e-3_h0.05_wDustgrowth} at $t=5000t_0$.
		Note that the stokes number in the region of $R/\rp \lesssim 1.1$ where the dust grains are almost depleted as $\sigmadust/\sigmagas < 10^{-4}$ is fixed to be 0.1 (see text).
		\label{fig:sizedist_a4e-3_h0.05_q1e-3}
		}
	\end{center}
\end{figure}
In Figure~\ref{fig:rings_q1e-3_a4e-3_h0.05_wDustgrowth}, we show the two-dimensional distribution of the surface density of the dust grains when the size evolution of the dust grains is considered ($\ufrag=30\mbox{m/s}$ is adopted).
The planet mass and the disk parameters are the same as these in Figure~\ref{fig:rings_q1e-3_a4e-3_h0.05_const_dustsize}.
When the size evolution of the dust grains is considered, the size of the dust grains is not constant, and it varies as described by Equation~(\ref{eq:size_representative}).
In Figure~\ref{fig:sizedist_a4e-3_h0.05_q1e-3}, we show the stokes number for the representative size of the dust grains at $t=5000t_0$, in the case of Figure~\ref{fig:rings_q1e-3_a4e-3_h0.05_wDustgrowth}.
As can be seen in Figure~\ref{fig:sizedist_a4e-3_h0.05_q1e-3}, the representative size of the dust grains is determined by the fragmentation in the outer disk.
In this case, the representative size is quite similar to the size of the grains in the  the case of the constant grain size.
Hence, the structure and the width of the dust ring are also similar to each other in the lower panel of Figure~\ref{fig:rings_q1e-3_a4e-3_h0.05_const_dustsize} and Figure~\ref{fig:rings_q1e-3_a4e-3_h0.05_wDustgrowth}.
At the outer edge of the gap ($R \sim 1.2R_0$), the dust-to-gas mass ratio sharply decreases and then $\dsizedrift$ becomes much smaller than $\dsizefrag$.
Because of it, the representative size is given by $\dsizedrift$ in this region.
Note that, as mentioned in Section~\ref{subsec:dust_size_evo}, the assumption of the coagulation--fragmentation equilibrium would not be valid and the model of \cite{Birnstiel_Klahr_Ercolano2012} may not be appropriate in this region.
Also note that in the region where almost no dust grains stay (i.e., $\sigmadust/\sigmagas < 10^{-4}$), the dust size is fixed to be $\st=0.1$ to avoid too small time step (see Section~\ref{subsec:dust_size_evo}).

Due to the rapid change of the representative size of the dust grains at the outer edge of the gap, the edge of the dust ring can be swung as in the right panel of Figure~\ref{fig:rings_q1e-3_a4e-3_h0.05_wDustgrowth}.
Such a swing may be related to the dust supply to the vicinity of the planet.
However, in this region, the size distribution of the dust grains could not reach the coagulation--fragmentation equilibrium.
To address this issue correctly, we have to consider more sophisticated model for the dust evolution.
% 
% \begin{figure}
% 	\begin{center}
% 		\resizebox{0.49\textwidth}{!}{\includegraphics{dgratio_dist_q1e-3_a4e-3.eps}}
% 		\caption{
% 		The azimuthally averaged value of the dust-to-gas mass ratio when $\mpl/\mstar=10^{-3}$, $\hgp/\rp=0.05$, and $\alpha=4\times 10^{-3}$ at $5000t_0$.
% 		\label{fig:dgratio_a4e-3_h0.05_q1e-3}
% 		}
% 	\end{center}
% \end{figure}
% Figure~\ref{fig:dgratio_a4e-3_h0.05_q1e-3} shows the azimuthally averaged values of the dust-to-gas mass ratio in the cases of Figures~\ref{fig:rings_q1e-3_a4e-3_h0.05_const_dustsize} and \ref{fig:rings_q1e-3_a4e-3_h0.05_wDustgrowth}.
% As can be seen in the figure (it also can be seen in the two-dimensional distributions of $\sigmadust$), when the dust feedback is considered, the very wide ring of the dust grains can be formed outside the planet orbit, though only the narrow ring is formed when the dust feedback is not considered.

\begin{figure}
	\begin{center}
		\resizebox{0.49\textwidth}{!}{\includegraphics{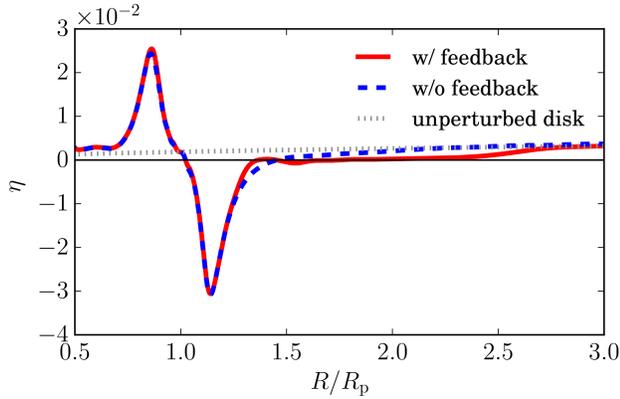}}
		\caption{
		The distribution of the azimuthally averaged value of $\eta$ in the case of Figure~\ref{fig:rings_q1e-3_a4e-3_h0.05_const_dustsize}.
		The solid and dashed lines denote the value of $\eta$ with and without the dust feedback, respectively.
		The gray dotted line denotes the value of $\eta$ in the unperturbed disk.
		\label{fig:eta_a4e-3_q1e-3}
		}
	\end{center}
\end{figure}
Figure~\ref{fig:eta_a4e-3_q1e-3} shows the azimuthally averaged value of $\eta$ (Equation~(\ref{eq:eta})) at $t=5000t_0$ in the case of Figure~\ref{fig:rings_q1e-3_a4e-3_h0.05_const_dustsize}.
A pressure bump is formed at the location where $\eta=0$ and $\partial \eta/\partial R > 0$.
The 'primary' pressure bump is formed at $R/\rp \simeq 1.4$ due to the gap formed by the planet in both the cases in which the dust feedback is considered and it is ignored.
The locations of the primary pressure bump are quite similar in both the cases.
It is remarkable that in the case in which the dust feedback is considered, the value of $\eta$ is very small as compared with that in the case in which the dust feedback is ignored, in the broad region within the dust ring.
Moreover, the secondary pressure bump is formed at $R/\rp = 1.6$, in the case in which the dust feedback is considered.
%The location of the second pressure bump roughly corresponds to the location of the maximum of the dust-to-gas mass ratio.
\begin{figure}
	\begin{center}
		\resizebox{0.49\textwidth}{!}{\includegraphics{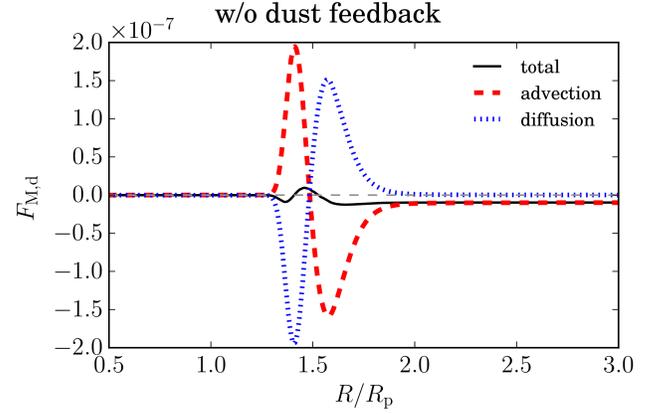}}
		\resizebox{0.49\textwidth}{!}{\includegraphics{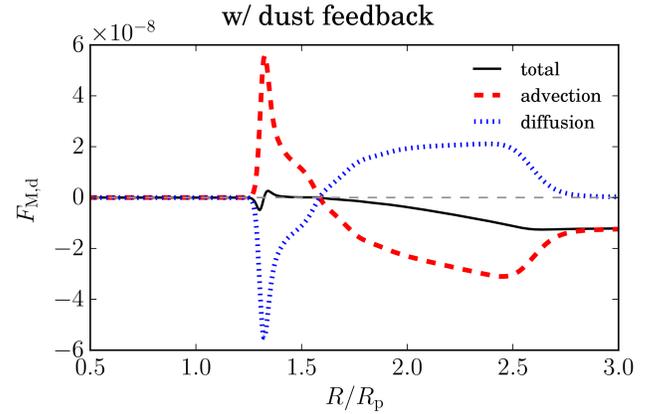}}
		\caption{
		The radial mass flux of the dust grains at $5000t_0$, in the case of Figure~\ref{fig:rings_q1e-3_a4e-3_h0.05_const_dustsize}.
		The dust feedback is ignored in the upper panel and it is considered in the lower panel.
		The dashed and dotted lines denote the mass flux of the dust grains due to the advection ($\vrdust \sigmadust$) and turbulent diffusion ($j_R$ given by Equation~(\ref{eq:j})), respectively, and the solid thin line is the sum of the advection and diffusion fluxes.
		\label{fig:dustmassflux_a4e-3_q1e-3}
		}
	\end{center}
\end{figure}
Figure~\ref{fig:dustmassflux_a4e-3_q1e-3} shows the radial mass flux of the dust grains at $t=5000t_0$ in the case of Figure~\ref{fig:rings_q1e-3_a4e-3_h0.05_const_dustsize}.
When the dust feedback is ignored, the mass flux due to the advection is so effective not to be canceled out by the diffusive mass flux outside the pressure bump.
Hence, the dust grains are accumulated within the narrow annulus near the pressure bump.
On the other hand, when the dust feedback is considered, the mass flux due to the advection is weakened because $\eta$ becomes small as shown in Figure~\ref{fig:eta_a4e-3_q1e-3}.
Because of it, the dust grains can diffuse from the dust ring and modify the  gas structure of the outer edge of the ring to let $\eta$ be small.
As a consequence, the dust grains can further diffuse from the edge of the ring and the width of the ring increases with time, due to the effects of the dust feedback.
Even when the growth of the dust grains is considered, the above picture does not change.
Note that although we adopt the model of the turbulent diffusion given by Equation~(\ref{eq:j}), this model may be appropriate when $\sigmadust \gtrsim \sigmagas$ and $\st \sim 1$.
The structure of the dust ring would depend on the model of the turbulent diffusion of the dust grains, which is discussed below again.

\subsubsection{Widths of the dust ring}
In this subsection, we consider the width of the dust ring.
First, we define the width of the ring as the radial width of the region where the azimuthally averaged value of the dust-to-gas mass ratio is larger than the threshold value ($\dgratioth$).
For convenience, we also define the radii of the inner and outer edges of the ring as $R_{\rm r,i}$ and $R_{\rm r,o}$, and $\rmean=(R_{\rm r,i}+R_{\rm r,o})/2$.

The width of the dust ring may be estimated as follows.
When the radial flow of the dust grains is almost halted, the mass of the dust ring would increase as the mass flux of the dust grains from the outside.
Hence, we can obtain $2\pi d(\rmean \wring \Sigma_{\rm d,mean})/dt \simeq -F_{\rm M,d}(R_{\rm out})$, where $\Sigma_{\rm d,mean}$ and $F_{\rm M,d}$ are the mean surface density of the dust grains in the ring and the mass flux of the dust grains from the outside, respectively.
If $\Sigma_{\rm d,mean} \sim \sigmagas \sim \Sigma_0$,
\begin{align}
\frac{d}{d(t/t_0)} \bfrac{\wring \rmean}{\rp^2}{} &= -\frac{F_{\rm M,d}(R_{\rm out})t_0}{2\pi \Sigma_0}.
\label{eq:dwringdt}
\end{align}
and $F_{\rm M,d}(R_{\rm out})$ is given by $2\pi R \sigmadust \vrdust$ at $R=R_{\rm out}$ and 
 $\vrdust$ is 
\begin{align}
\vrdust(R) &= \frac{2(\dsize/\dsizeo) \eta_{\rm p}}{1+(\dsize/\dsizeo)^{2}(R/\rp)^{2s}} \eta_{\rm p}\bfrac{R}{\rp}{2f+3/2}.
\label{eq:vrad_dust}
\end{align}
Hence, 
\begin{align}
f_{\rm M,d} &\equiv -\frac{F_{\rm M,d}(R_{\rm out})t_0}{2\pi \Sigma_0}\nonumber \\
&= \frac{4\pi(\dsize/\dsizeo) \eta_{\rm p} (\sigmadust/\sigmagas)_{\rm out}}{1+(\dsize/\dsizeo)^{2}(R_{\rm out}/\rp)^{2s}} \bfrac{R_{\rm out}}{\rp}{2f+3/2}.
\label{eq:fm_out}
\end{align}
Note that the value of $\wring \rmean$ indicates the area of the ring region.
Hence the value of $\wring \rmean$ is described by a linear function of the time as
\begin{align}
\frac{\wring \rmean}{\rp^2} &= \frac{f_{\rm M,d}}{t_0} \left(t - t_r \right),
\label{eq:linear_func}
\end{align}
where in $t>t_r$, the ring of the dust grains could be formed.  

\begin{figure}
	\begin{center}
		\resizebox{0.49\textwidth}{!}{\includegraphics{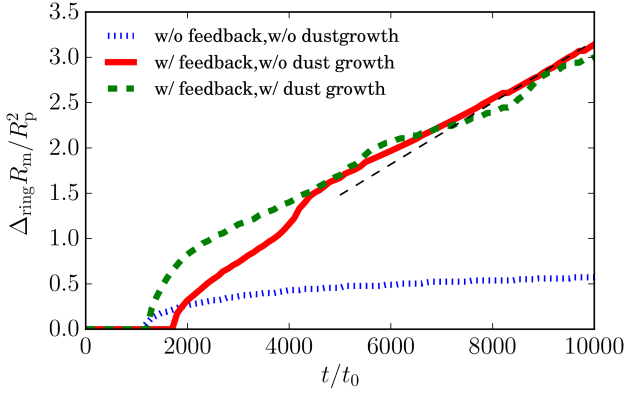}}
		\caption{
		The time variations of the area of the dust ring ($\wring \rmean$) in the same cases as Figures~\ref{fig:rings_q1e-3_a4e-3_h0.05_const_dustsize} and \ref{fig:rings_q1e-3_a4e-3_h0.05_wDustgrowth}.
		The widths are measured by the radial width of the region where the dust-to-gas mass ratio is larger than $\dgratioth=0.5$.
		The dashed thin line denote Equation~(\ref{eq:linear_func}) with $f_{\rm M,d}(R_{\rm out}) = 3.4 \times 10^{-4}$ and $t_{r} = 882t_0$.
		\label{fig:tevo_ring_a4e-3_q1e-3}
		}
	\end{center}
\end{figure}
Figure~\ref{fig:tevo_ring_a4e-3_q1e-3} shows the time variations of the area of the dust ring ($\wring \rmean$) measured by $\dgratioth=0.5$, in the cases of Figures~\ref{fig:rings_q1e-3_a4e-3_h0.05_const_dustsize} and \ref{fig:rings_q1e-3_a4e-3_h0.05_wDustgrowth}.
If the dust feedback is not considered, the area of the dust ring does not increase after $t \sim 1000t_0$.
On the other hand, when the dust feedback is considered, the area of the dust ring increases with time.
When the size of the dust grains is constant throughout the disk, the value of $f_{\rm M,d}(R_{\rm out})$ is given by $3.4\times 10^{-4}$, because $\dsize=0.1\dsizeo$, $\eta_{\rm p}=2\times 10^{-3}$, $(\sigmadust/\sigmagas)_{\rm out}=0.01$, and $R_{\rm out}=4\rp$.
As can be seen in Figure~\ref{fig:tevo_ring_a4e-3_q1e-3}, the linear function of Equation~(\ref{eq:linear_func}) with $f_{\rm M,d}=3.4\times 10^{-4}$ can reproduce the time variation of the area of the dust ring in the case with the constant size of the dust grains, after $t=5000t_0$ (in this case, $t_r=882t_0$).
Even when the growth of the dust grains is considered, the time-variation of the dust ring is quite similar to that in the case of the constant-sized dust grains, because the representative size is similar to that given by $\st \sim 0.1$ as shown in Figure~\ref{fig:sizedist_a4e-3_h0.05_q1e-3}.
Hence, in this case, the time variation of the dust ring is also reproduced well by the same function.

\begin{figure}
	\begin{center}
		\resizebox{0.49\textwidth}{!}{\includegraphics{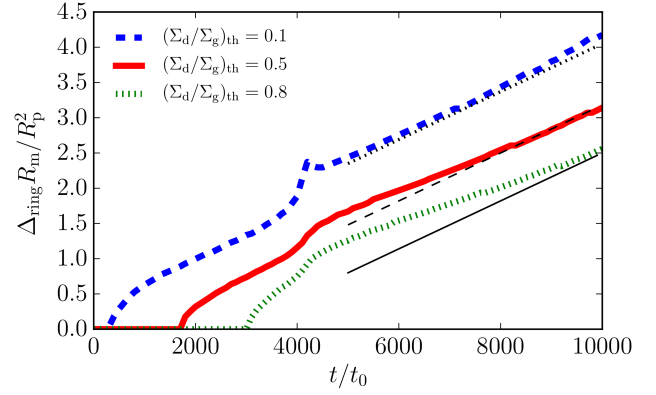}}
		\caption{
		The time variation of the area of the dust ring whose width is measured by  $\dgratioth=0.1$, $0.5$, and $0.8$, in the case with the constant size of the dust grains.
		The planet mass and the disk parameter are the same as these in Figure~\ref{fig:rings_q1e-3_a4e-3_h0.05_const_dustsize}.
		The dashed thin line is the same as that in Figure~\ref{fig:tevo_ring_a4e-3_q1e-3}.
		The solid and dotted thin lines denote Equation~(\ref{eq:linear_func}) with $f_{\rm M,d}(R_{\rm out}) = 3.4 \times 10^{-4}$ and $t_{r} = 2649t_0$ and $t_{r} = 1911 t_0$, respectively.
		\label{fig:tevo_ring_a4e-3_q1e-3_vardgratio}
		}
	\end{center}
\end{figure}
Figure~\ref{fig:tevo_ring_a4e-3_q1e-3_vardgratio} shows the time variation of the area of the dust ring measured by $\dgratioth=0.1$, $0.5$, and $0.8$ in the case with the constant size of the dust grains (lower panel of Figure~\ref{fig:rings_q1e-3_a4e-3_h0.05_const_dustsize}).
When the ring is measured by $\dgratioth=0.1$ and $0.8$, after $t\simeq 5000t_0$, the time variations of the area of the dust ring can be reasonably reproduced by Equation~(\ref{eq:linear_func}) with the same value of $f_{\rm M,d}(R_{\rm out})$ in the case of $\dgratioth=0.5$, whereas the value of $t_r$ is different in each case.

\subsection{Dependence on the mass of the planet} \label{subsec:ring_various_mass}
\begin{figure}
	\begin{center}
		\resizebox{0.49\textwidth}{!}{\includegraphics{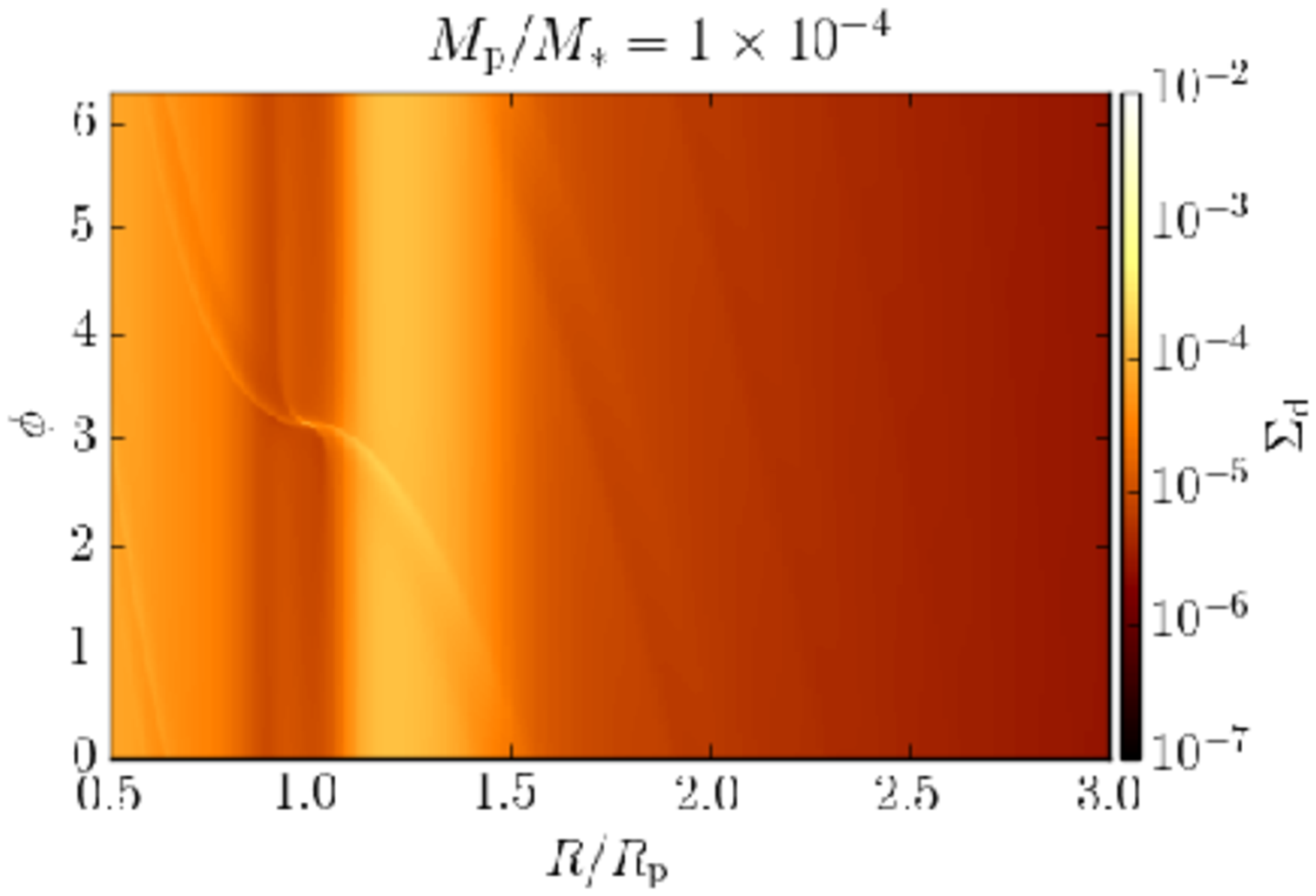}}
		\resizebox{0.49\textwidth}{!}{\includegraphics{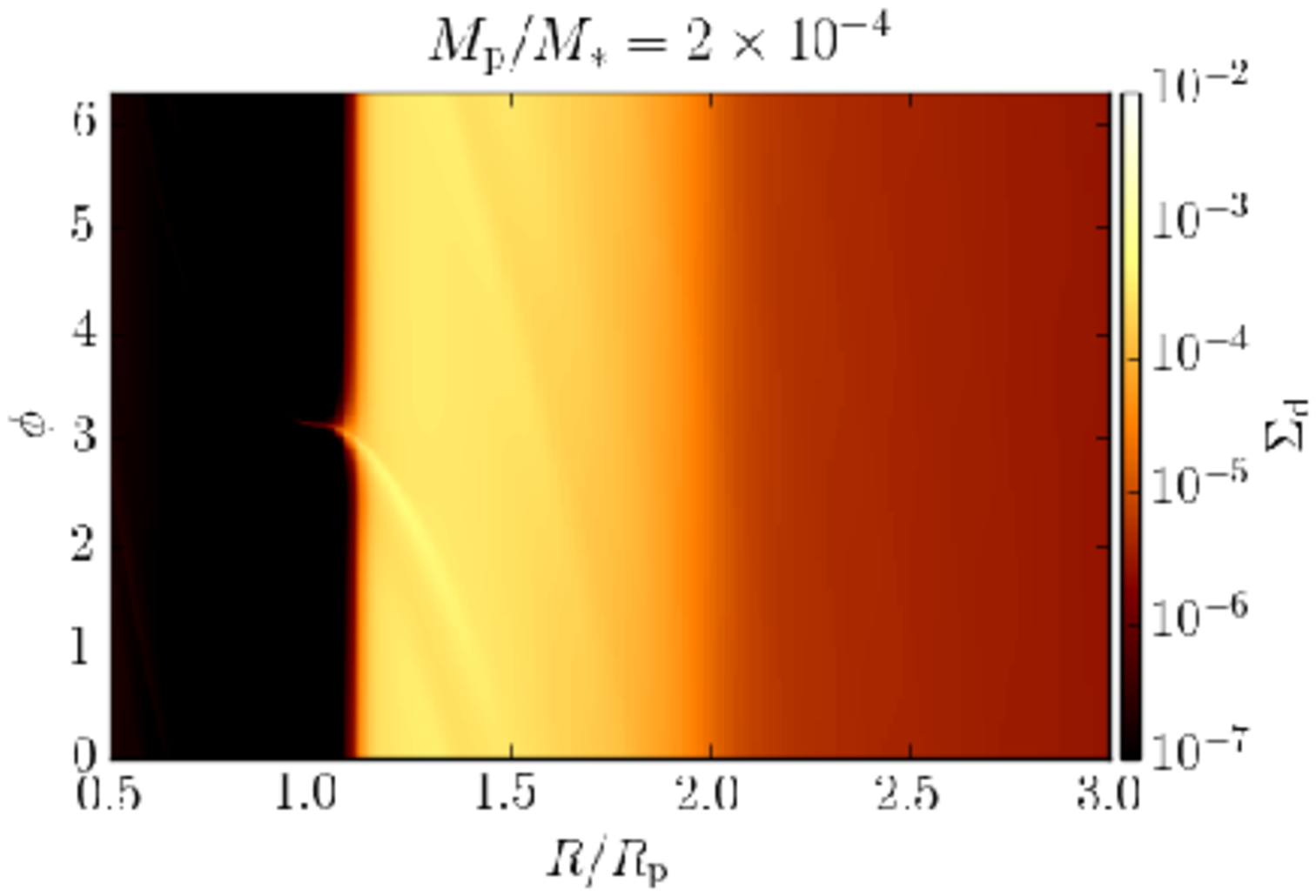}}
		\resizebox{0.49\textwidth}{!}{\includegraphics{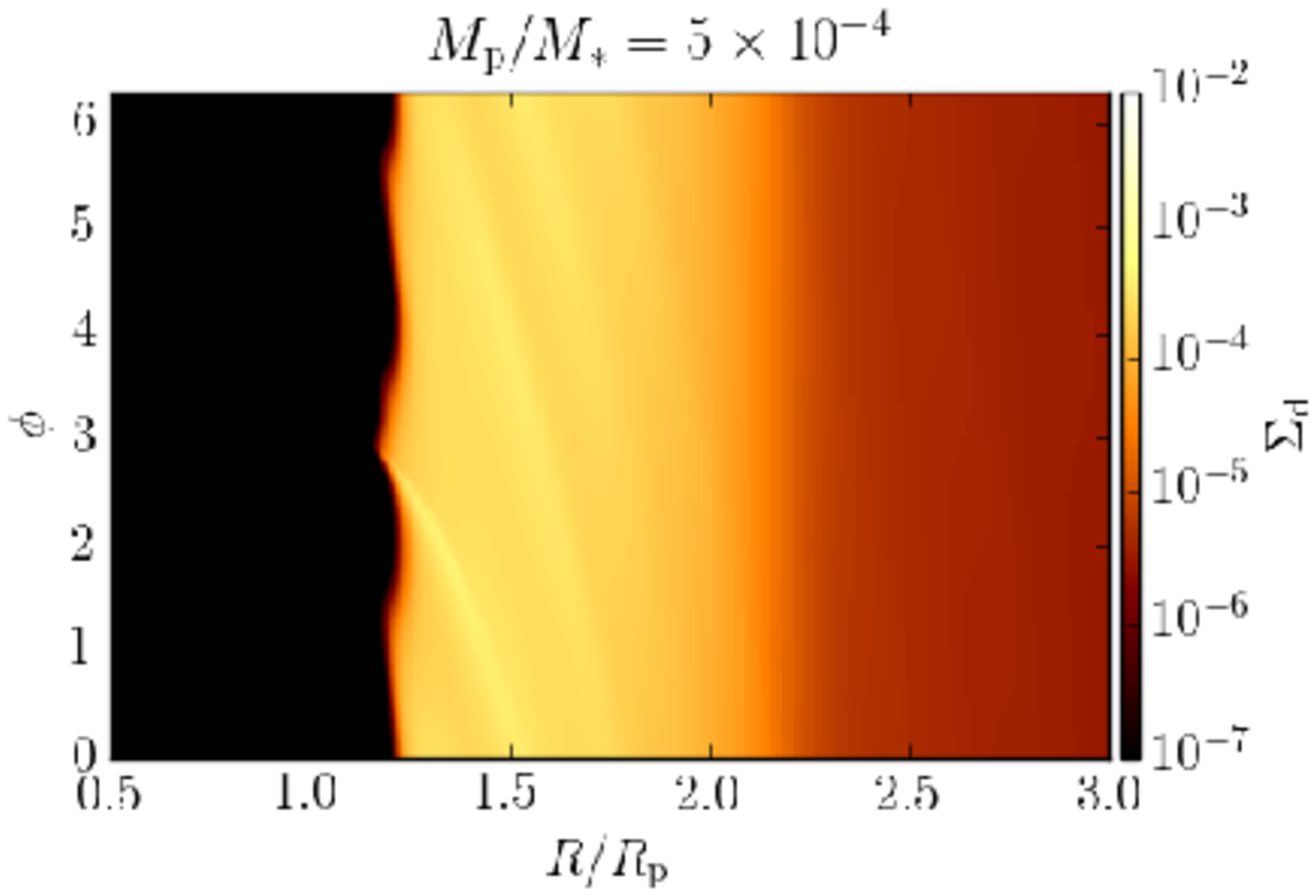}}
		\caption{
		The two-dimensional distributions of the surface density of the dust grains at $t=5000t_0$, when $\mpl/\mstar=1\times 10^{-4}$, $2\times 10^{-4}$, and $5\times 10^{-4}$ from the top panel.
		In the figure, $\alpha=4\times 10^{-3}$ and $\hgp/\rp =0.05$.
		\label{fig:dust_dist_a4e-3_qvar}
		}
	\end{center}
\end{figure}
\begin{figure}
	\begin{center}
		\resizebox{0.49\textwidth}{!}{\includegraphics{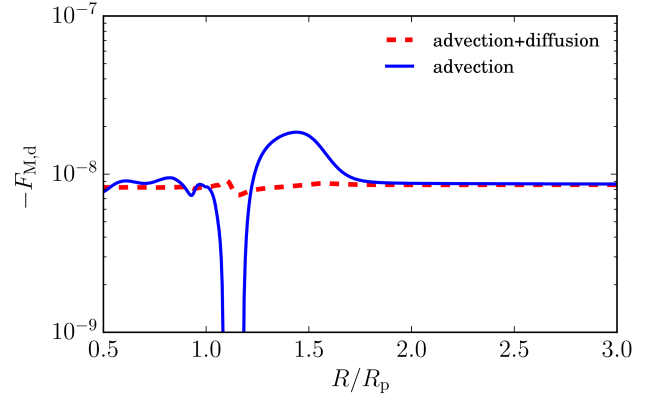}}
		\caption{
		The radial mass flux in the case of $\mpl/\mstar=1\times 10^{-4}$ and $\alpha=4\times 10^{-3}$.
		The dashed line shows the radial mass flux due to advection ($\vrdust \sigmadust$) and the solid line shows the radial mass flux which is the sum of the advection flux and diffusion flux (see Equation~(\ref{eq:dust_mass_flux})).
		\label{fig:dustmassflux_a4e-3_q1e-4}
		}
	\end{center}
\end{figure}
Figure~\ref{fig:dust_dist_a4e-3_qvar} shows the two-dimensional distributions of the surface density of the dust grains for various planet masses, in the case with the constant size of the dust grains.
In the cases of $\mpl/\mstar=1\times 10^{-4}$, the inner disks of the dust grains are not depleted, though the relatively deep gap of the dust grains is formed.
In this case, one can find that the dust grains penetrate into the gap region around the opposite site from the planet.
Figure~\ref{fig:dustmassflux_a4e-3_q1e-4} shows the radial mass flux due to the advection and the sum of the advection and diffusion fluxes, in the case of $\mpl/\mstar=1\times 10^{-4}$.
As can be seen in the figure, in the outer disk close to the planet, the dust grains are carried by the diffusive mass flux.
Hence in this case, the structure of the dust grains is kept the steady state by the diffusion.
On the other hand, when $\mpl/\mstar = 2\times 10^{-4}$ and $5\times 10^{-4}$, the surface density of the dust grains in the inner disk is significantly small and the inner disk of the dust grains is almost depleted.
In these cases, the radial mass flow of the dust grains across the gap is almost halted.
Moreover, only in these cases, the wide ring of the dust grains can be formed.

The minimum mass of the planet to form the dust ring should correspond to the pebble-isolation mass \citep[e.g.,][]{Morbidelli_Nesvorny2012,Lambrechts_Johansen_Morbidelli2014,Bitsch_Morbidelli_Johansen_Lega_Lambrechts_Crida2018}.
Recently \cite{Bitsch_Morbidelli_Johansen_Lega_Lambrechts_Crida2018} have obtained a formula of the pebble-isolation mass, by carrying out three-dimensional hydrodynamic simulations.
The pebble-isolation mass given by their formula (Equation~(26) of that paper) is about $1.3\times 10^{-4} \mstar$ when $\alpha=4\times 10^{-3}$ and $\hgp/\rp =0.05$.
Their pebble-isolation mass reasonably agrees with our minimum mass of the planet to form the broad ring of the dust grains.

% \begin{figure}
% 	\begin{center}
% 		\resizebox{0.49\textwidth}{!}{\includegraphics{dustmassflux_a1e-2.eps}}
% 		\caption{
% 		The radial mass fluxes of the dust grains at $t=10000t_0$ in the cases with $\mpl/\mstar=1\times 10^{-4}$ -- $5\times 10^{-4}$.
% 		The value of $\alpha$ is set to be $10^{-2}$ and $\hgp/\rp =0.05$ in the figure.
% 		\label{fig:dustmassflux_a1e-2_qvar}
% 		}
% 	\end{center}
% \end{figure}
% In Figure~\ref{fig:dustmassflux_a1e-2_qvar}, we show the radial mass fluxes of the dust grains for various planet masses, in the case with $\alpha=10^{-2}$.
% When the planet mass is smaller than $3\times 10^{-4}\mstar$, the radial mass flux of the dust grains is almost constant throughout the disk, though it slightly varies in the vicinity of the planet.
% Hence, in this case, the structure of the dust disk reaches the steady state.
% On the other hand, when $\mpl/\mstar=5\times 10^{-4}$, the radial mass flux of the dust grains significantly decreases due to the planet-induced gap.
% In this case, almost no dust grains can pass though the gap region and therefore, the inner disk of the dust grains is depleted quickly and the wide ring of the dust grains is formed.

\begin{figure}
	\begin{center}
		\resizebox{0.49\textwidth}{!}{\includegraphics{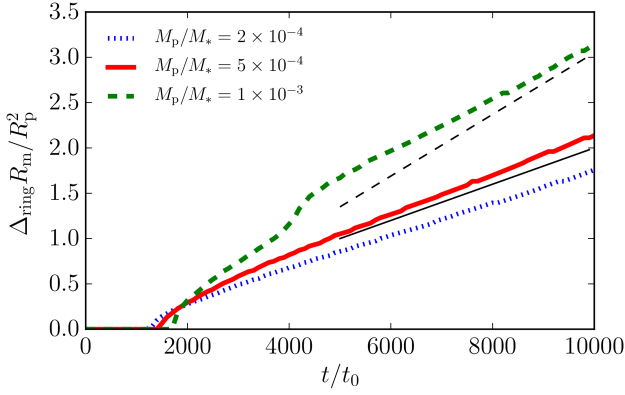}}
		\caption{
		The same as Figure~\ref{fig:tevo_ring_a4e-3_q1e-3}, but for various masses of the planet.
		The dashed thin line and solid thin line denote Equation~(\ref{eq:linear_func}) with $f_{\rm M,d}(R_{\rm out}) = 3.4 \times 10^{-4}$ and $t_{r} = 1029 t_0$ and that with $f_{\rm M,d}(R_{\rm out}) = 2.0 \times 10^{-4}$ and $t_{r} = 0$, respectively.
		\label{fig:tevo_widths_qvar_a4e-3}
		}
	\end{center}
\end{figure}
Figure~\ref{fig:tevo_widths_qvar_a4e-3} shows the time variation of the area of the dust rings in the cases of various planet masses.
When $\mpl/\mstar=2\times 10^{-4}$ and $\mpl/\mstar=5\times 10^{-4}$, the width of the dust ring increases with time, as in the case of $\mpl/\mstar=10^{-3}$.
However, the increase rates of the dust ring area in the cases with $\mpl/\mstar= 2\times 10^{-4}$ and $\mpl/\mstar = 5\times 10^{-4}$ are slightly smaller than that in the case with $\mpl/\mstar=10^{-3}$.
This discrepancy could be originated from the different from the surface density of the gas within the dust ring.
As shown in Appendix~\ref{sec:gas_disk_structures}, the surface density of the gas within the dust ring is slightly larger when the planet mass is smaller (or the gap is shallower).
Hence, since the dust-to-gas mass ratio is approximately unity within the dust ring, the surface density of the dust grains is larger with the smaller mass of the planet.
Because of it, the increase rate of the dust ring is smaller when the planet mass is relatively small.
Nonetheless, we can roughly estimate the width of the dust ring by Equation~(\ref{eq:dwringdt}) because the discrepancy from Equation~(\ref{eq:dwringdt}) is not more than a factor of two.

\subsection{In the case of a low viscosity} \label{subsec:ring_low_viscosity}
\begin{figure}
	\begin{center}
		\resizebox{0.49\textwidth}{!}{\includegraphics{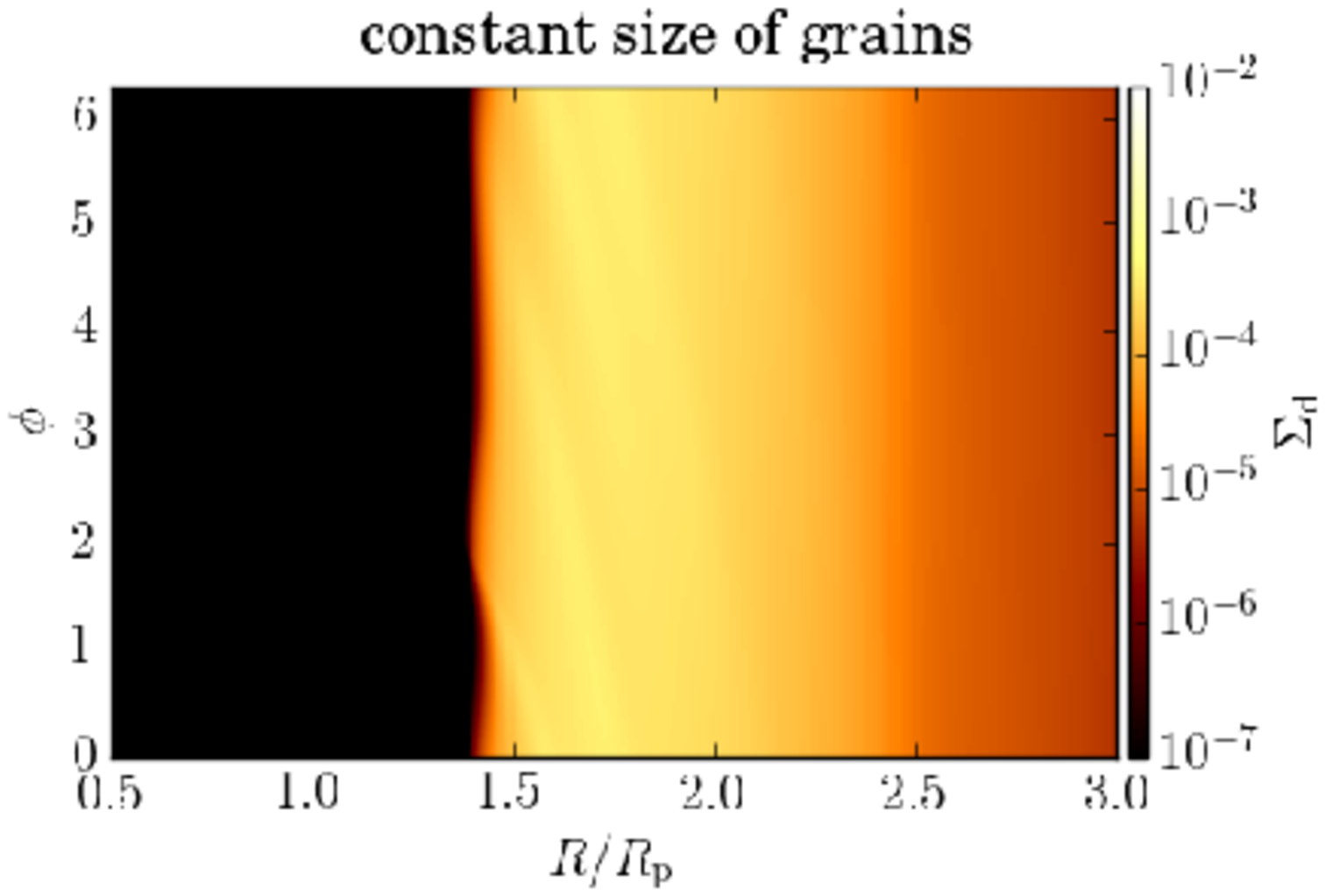}}
		\resizebox{0.49\textwidth}{!}{\includegraphics{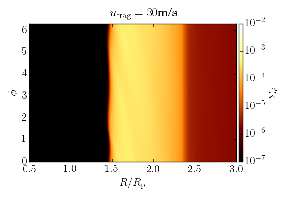}}
		\resizebox{0.49\textwidth}{!}{\includegraphics{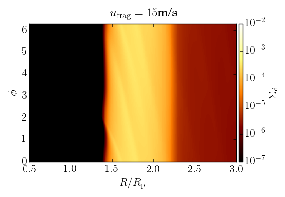}}
		\caption{
		The two-dimensional distributions of the dust surface density at $t=5000t_0$ when $\mpl/\mstar=10^{-3}$, $\hgp/\rp=0.05$ and $\alpha=10^{-3}$.
		In the top panel, the dust size is constant during the simulation, while in the middle and lower panels, the size evolution of the dust grains is considered with $\ufrag=30\mbox{m/s}$ and $\ufrag=15\mbox{m/s}$, respectively.
		\label{fig:rings_a1e-3_q1e-3}
		}
	\end{center}
\end{figure}
\begin{figure}
	\begin{center}
		\resizebox{0.49\textwidth}{!}{\includegraphics{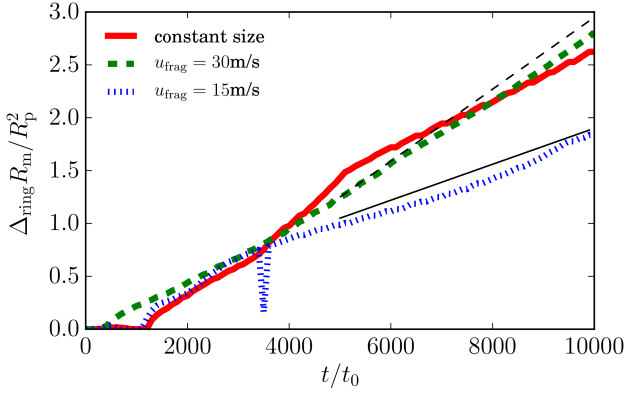}}
		\caption{
		The time variations of the area of the dust ring in the case of Figure~\ref{fig:rings_a1e-3_q1e-3}.
		The dashed thin line and the solid line denote Equation~(\ref{eq:linear_func}) with $f_{\rm M,d}(R_{\rm out}) = 3.4\times 10^{-4}$ and $t_{r} = 1323$ and that with $f_{\rm M,d}(R_{\rm out}) = 1.7 \times 10^{-4}$ and $t_{r} = -1176 t_0$, respectively.
		\label{fig:tevo_widths_a1e-3_q1e-3}
		}
	\end{center}
\end{figure}
According to Equation~(\ref{eq:fm_out}), the increase rate of the ring area is independent of the viscosity.
Figure~\ref{fig:rings_a1e-3_q1e-3} shows the two-dimensional distributions of the dust surface density at $t=5000t_0$ when $\alpha=10^{-3}$ and $\hgp/\rp=0.05$.
As in the case of $\alpha=4\times 10^{-3}$ shown in Figures~\ref{fig:rings_q1e-3_a4e-3_h0.05_const_dustsize} and \ref{fig:rings_q1e-3_a4e-3_h0.05_wDustgrowth}, the broad dust ring is formed in the outer disk in both the cases where the size evolution of the dust grains is ignored and it is considered.
In this case, since the collision velocity due to the turbulence is smaller than that when $\alpha=4\times 10^{-3}$, the broad dust ring can be formed even if $\ufrag=15\mbox{m/s}$.
When $\ufrag$ is small, the representative size of the dust grains becomes also small (see Equation~(\ref{eq:size_frag})).
Hence, the increase rate of the ring area decreases with a smaller $\ufrag$.
Figure~\ref{fig:tevo_widths_a1e-3_q1e-3} shows the time variations of the area of the dust ring in the case shown in Figure~\ref{fig:rings_a1e-3_q1e-3}.
In both the case in which the dust size is constant and the case in which the dust growth is considered with $\ufrag=30\mbox{m/s}$, the increases rate of the ring area is quite similar to each other, and it is also similar to that in the case of $\alpha=4\times 10^{-3}$ and $\mpl/\mstar=10^{-3}$.
When $\ufrag=15\mbox{m/s}$, the increase rate of the ring area is about 0.5 times that in the case of $\ufrag=30\mbox{m/s}$, as mentioned above.

\begin{figure}
	\begin{center}
		\resizebox{0.49\textwidth}{!}{\includegraphics{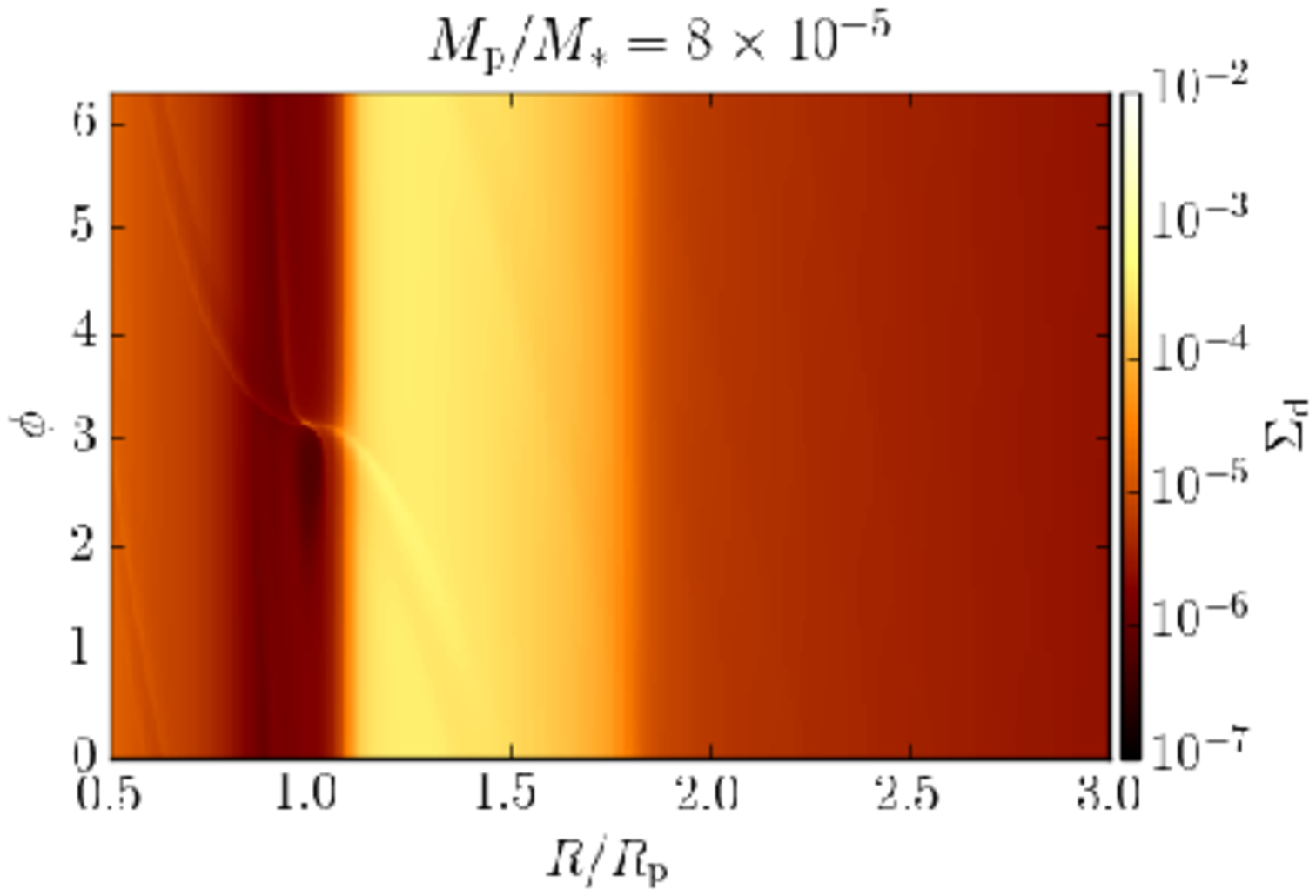}}
		\resizebox{0.49\textwidth}{!}{\includegraphics{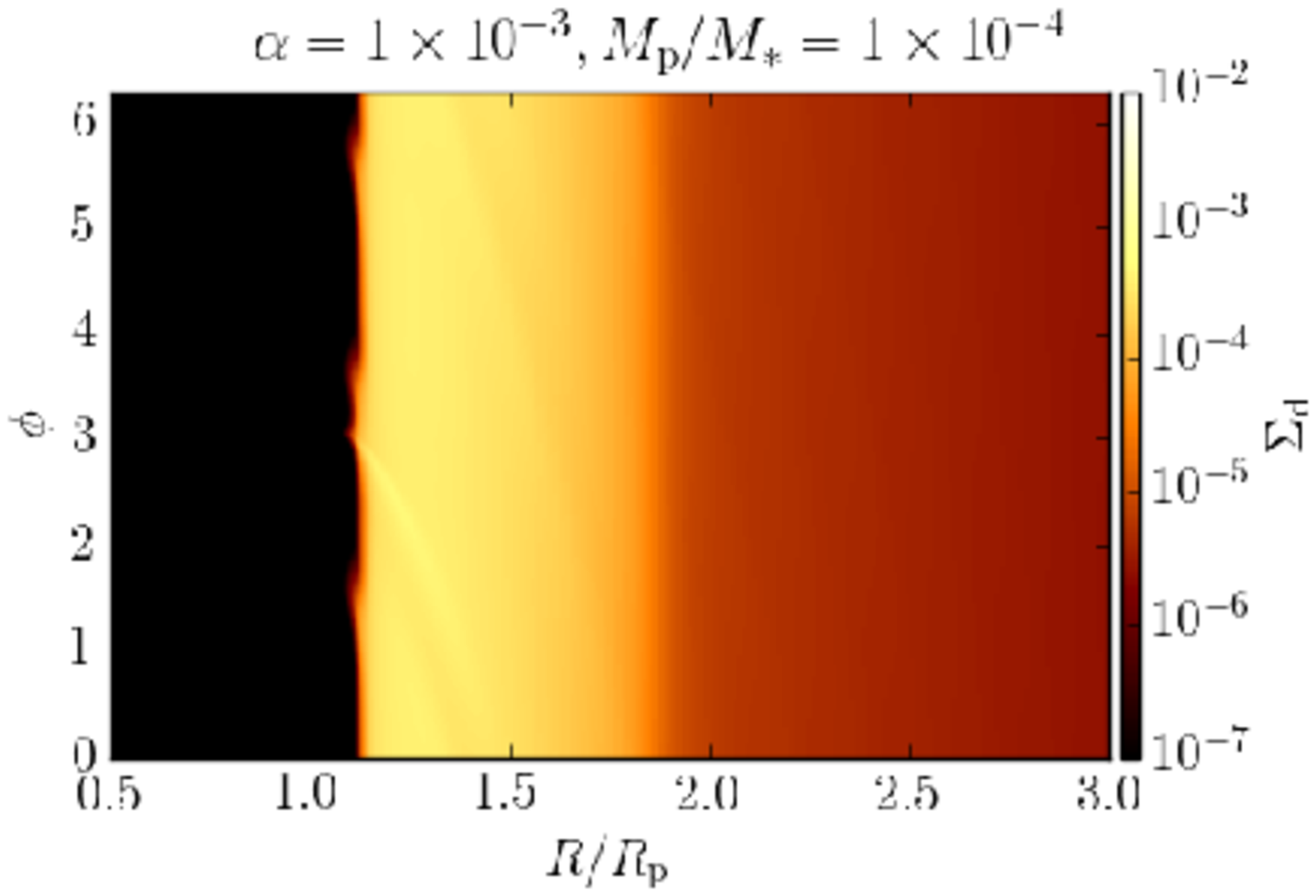}}
		\caption{
		The two-dimensional distributions of the dust surface density at $t=5000t_0$ when $\hgp/\rp=0.05$ and $\alpha=10^{-3}$.
		The planet masses are $\mpl/\mstar=8\times 10^{-5}$ in the top panel and $\mpl/\mstar=10^{-4}$ in the lower panel, respectively.
		\label{fig:dens2d_a1e-3}
		}
	\end{center}
\end{figure}
When the viscosity is small, the minimum mass of the planet to form the broad dust ring is also small, since even small planet can form the gas gap.
Figure~\ref{fig:dens2d_a1e-3} illustrates the two-dimensional distributions of the dust surface density at $t=5000t_0$ when $\alpha=10^{-3}$ and the constant sized dust grains are adopted.
As can be seen in the figure, when $\mpl/\mstar=8\times 10^{-5}$, the inner disk of the dust grains still remain and the dust grains can pass through the gap.
In this case, a relatively broad ring is formed at $t=5000t_0$, but it does not increases anymore.
While when $\mpl/\mstar=10^{-4}$, the inner disk of the dust grains is almost depleted and the dust grains cannot pass through the gap.
In this case, the area of the dust ring increases with time.
Hence, in this case, we can consider that the minimum mass of the planet to form the broad dust ring is $\mpl/\mstar \simeq 10^{-4}$.
This minimum mass also reasonably agrees with the pebble-isolation mass given by \cite{Bitsch_Morbidelli_Johansen_Lega_Lambrechts_Crida2018}.

It should be noted that as shown in Section~\ref{subsec:ring_jupiter}, the dust ring enlarges by that the diffusing dust grains from the edge of the ring modify the gas structure at the edge of the ring.
Hence, the growth timescale of the dust ring becomes longer as the turbulent diffusivity of the dust grains becomes ineffective.
As shown above, the dust ring can grow wider during the simulation time (i.e. $10000t_0$) in our parameter range as $\alpha \geq 10^{-3}$.
However, if the turbulent diffusivity of the dust grains is extremely low, this timescale would be significantly longer.
In addition, in such a low viscous case, the collision velocity induced by the radial drift could be superiority to that induced by the turbulence.
Hence, we need to adapt more sophisticated model of the dust grains in this case.

\section{Discussion} \label{sec:discussion}
\subsection{Formation of the planetesimals via the streaming instability} \label{subsec:ef_formation_planetesimals}
For simplicity, we do not include the formation of planetesimals in our simulations.
In the region in which the dust grains are accumulated such as the dust ring, however, the planetesimals can be effectively formed via the streaming instability \citep[e.g.,][]{Youdin_Goodman2005,Youdin_Johansen2007,Johansen_Youdin2007,Drazkowska_Dullemond2014,Carrera_Johansen_Davies2015,Auffinger_Laibe2018}.
If most of the dust grains are instantaneously converted to the planetesimals by the streaming instability, the dust ring could not grow anymore.
However, not all the dust grains may be converted to the planetesimal \citep{Johansen_Youdin_Lithwick2012,Simon_Armitage_Li_Youdin2016}.
If the efficiency for the formation of the planetesimals is moderate, it is possible that the dust ring can grow up while the planetesimals are formed.
Moreover, the value of $\eta$ becomes very small within the dust ring as shown in Figure~\ref{fig:eta_a4e-3_q1e-3}, which affects the spatial scale and the growth timescale of the streaming instability \citep{Youdin_Goodman2005,Youdin_Johansen2007}.
With a smaller value of $\eta$, the spatial scale of the streaming instability becomes smaller and the growth timescale is longer.
The small value of $\eta$ could prevent an effective formation of the planetesimals by the streaming instability.
When the timescale of the planetesimals is comparable with or longer than the growth timescale of the dust ring, the broad dust ring could be formed.

\subsection{Implication for planet formation} \label{subsec:planet_formation}
Planetesimals can be produced within the ring with the high dust density formed outside the planet-induced gap, as discussed by previous studies \citep[e.g.,][]{Kobayashi_Ormel_Ida2012,Pinilla_Ovelar_Ataiee_Benisty_Birnstiel_Dishoeck_Min2015}.
In the previous section, we have shown that the very wide ring of the dust grains can be formed outside the planet-induced gap when the dust feedback is considered.
This fact implies that the planetesimal formation can be triggered in a wide region outside the planet-induced gap.

The maximum area of the dust ring can be constrained by the total mass of the dust grains within the protoplanetary disk.
If the radius of the protoplanetary disk holding the dust grains is given by $R_{\rm d,out}$ and the initial dust-to-gas mass ratio is $(\sigmadust/\sigmagas)_{\rm init}$ throughout the disk, the total dust mass can be roughly estimated by $(\sigmadust/\sigmagas)_{\rm init} \Sigma_0 R_{\rm d,out}^2$.
Considering $\sigmadust/\sigmagas \sim 1$ within the dust ring, we can roughly estimate the maximum area of the ring as
\begin{align}
\left(\frac{\wring \rmean}{\rp^2} \right)_{\max} &= \bfrac{\sigmadust}{\sigmagas}{}_{\rm init} \bfrac{R_{\rm d,out}}{\rp}{2}.
\label{eq:maximum_area_ring}
\end{align}
Assuming $R_{\rm r,i} = \rp$, we obtain
\begin{align}
\wring &= \rp \left(\sqrt{1+2\bfrac{\sigmadust}{\sigmagas}{}_{\rm init} \bfrac{R_{\rm d,out}}{\rp}{2}} -1 \right).
\label{eq:maximum_width_ring}
\end{align}
When $(\sigmadust/\sigmagas)_{\rm init} =0.01$, $R_{\rm d, out}= 100 \au$ and $\rp = 2 \au$, we can estimate $\wring \simeq 10\au$.
Similarly, if in a compact disk as $R_{\rm d,out}=10\au$ and $\rp=1\au$, we can estimate $\wring \simeq 1\au$.

Once the planetesimal forms, its mass increases by capturing dust grains within its feeding zone until its mass reaches the core-isolation mass \citep[e.g.,][]{Kokubo_Ida1995,Kokubo_Ida2000}.
The core-isolation mass ($M_{\rm pl,iso}$) can be estimated by $2\pi \sigmadust \rp \Delta_{\rm f}$, where $\Delta_{\rm f}$ is a width of the feeding zone.
The width of the feeding zone is approximately given by $10 \rp (M_{\rm pl,iso}/(3\mstar))^{1/3}$ \citep{Kokubo_Ida1995}.
The core-isolation mass and the width of its feeding zone can be obtained by
\begin{align}
M_{\rm pl,iso} &= 3.6 \bfrac{\sigmadust}{100\sigmaunit}{3/2} \bfrac{\mstar}{1M_{\odot}}{-1/2}  \bfrac{\rp}{1\au}{3} M_{\oplus},
\label{eq:isolation_mass} \\
\Delta_{\rm f} &= 0.15 \bfrac{\sigmadust}{100 \sigmaunit}{1/2} \bfrac{\mstar}{1M_{\odot}}{-1/2}\bfrac{\rp}{1\au}{2}\ \au
\label{eq:width_feeding_zone}.
\end{align}

First we consider the disk around the sun-like star ($M_{\rm disk} \sim 10^{-2} M_{\odot}$, $R_{\rm d,out} \sim 100 \au$, and $\mstar\sim 1M_{\odot}$).
When $\rp=10\au$ and $\sigmadust \simeq \sigmagas = 10 \sigmaunit$, the maximum width of the dust ring can be estimated by $7 \au$, which is comparable with the width of the feeding zone 
 ($\simeq 5\au$).
In this case, a massive core ($M_{\rm pl,iso} \simeq 100M_{\oplus}$) can be formed in the dust ring.
However, since this core-isolation mass is much larger than the critical core mass of the runaway gas accretion (typically $\sim 10M_{\oplus}$) \citep[e.g.,][]{Mizuno1980,Kanagawa_Fujimoto2013}, most of the dust grains would accrete onto the gaseous envelope after the runaway gas accretion, rather than the core.
Hence, the solid core of the giant planet may not be so massive in this case, instead of a metal-rich gas envelope.
On the other hand, when the ring is formed in the inner region of the disk, a giant planet holding a massive solid core might be formed.
When we consider $\rp=2\au$ and $\sigmadust \simeq \sigmagas = 200 \sigmaunit$, $\Delta_{\rm f}$ is about $1\au$.
Since the maximum width of the dust ring estimated by Equation~(\ref{eq:maximum_width_ring}) is about $10\au$, which is much wider than the width of the feeding zone.
In this case, as shown by \cite{Ikoma_Guillot_Genda_Tanigawa_Ida2006}, since the critical core mass for the runaway gas accretion increases due to high core accretion rate, the solid core can grow up to be the isolation mass ($M_{\rm pl,iso} \simeq 80 M_{\oplus}$) before the runaway gas accretion.
After strong gravitational scattering and giant impact events, a gas giant planet holding the massive solid core is possible to be formed.
The observations have found compact Hot Jupiters (giant planet with a very close orbit) with the massive solid core as $\mpl \sim 50M_{\oplus}$.
For instance, HD~149026b is suggested to hold the core of $50$--$80M_{\oplus}$ \citep{Sato2005,Fortney_Saumon_Marley_Lodders_Freedman2006,Ikoma_Guillot_Genda_Tanigawa_Ida2006}, and WASP-59b is also suggested to have the massive dens core \citep{Hebrard2013}.
Such a giant planet holding a massive solid core might be formed within the ring of the dust grains.
% The formation of giant planets holding such a massive solid core is difficult to understand both in core-accretion scenario and gravitational instability scenario.
% However, as \cite{Ikoma_Guillot_Genda_Tanigawa_Ida2006} have shown, it is possible in the inner region of the disk (i.e., $0.3 \au \lesssim R \lesssim 2 \au$) with the high density of the dust grains around the sun-like star (see Figure 6 in that paper).

Next, let us consider the dust ring in the disk around the dwarf star.
Recently seven Earth-sized planets have been found around the dwarf star named TRAPPIST-1 ($\mstar \simeq 0.08 M_{\odot}$) \citep{Gillon2016,Gillon2017}.
All seven planets orbit within $0.1\au$ from the host star and all planets have similar masses to Earth.
Here we consider the formation of the planetary system around the small mass star like TRAPPIST-1.
Hence, we consider $\mstar=0.1M_{\odot}$ and the compact and less massive disk ($R_{\rm d,out}=10\au$ and $M_{\rm disk} \sim 10^{-3} M_{\odot}$).
Here we consider $\rp=0.5\au$.
The surface density of disk gas around $0.5\au$ can be considered as $100 \mbox{ g/cm}^2$.
In the dust ring, the surface density of the dust grains is the same level of $\sigmagas$.
According to Equation~(\ref{eq:maximum_area_ring}), in this case, the width of the dust ring can reach up to $1\au$.
The width of the feeding zone is about $0.12\au$ as can be seen in Equation~(\ref{eq:width_feeding_zone}), and the mass of the solid core is about $1.4M_{\oplus}$ (Equation~(\ref{eq:isolation_mass})).
Hence, we can consider that several (up to eight) Earth-sized planets can be formed in the dust ring, in this case.
In addition, since the mass of the central star is small, even when its mass is small, the planet can form the dust ring in the inner region of the planet.
According to the formula given by \cite{Bitsch_Morbidelli_Johansen_Lega_Lambrechts_Crida2018}, when $\hgp/\rp=0.03$ and $\alpha=10^{-3}$, the pebble-isolation mass is about $\mpl/\mstar=1.6\times 10^{-5}$, which corresponds to $0.5M_{\oplus}$ when $\mstar=0.1M_{\odot}$.
The above discussion implies that once the small planet is formed in the small disk hosted by the dwarf star, as a consequence, multiple planetary system of the Earth-sized planets is formed.
The multiple system of Earth-sized planets may be common around the dwarf star like TRAPPIST-1.

\subsection{Implication for observation} \label{subsec:observation}
As shown in Figure~\ref{fig:sizedist_a4e-3_h0.05_q1e-3}, the size of the dust grains trapped into the dust ring is approximately given by $\dsize=0.1\dsizeo$, which corresponds to $\sim 3 \mbox{ cm}$ in the fiducial disk as $\sigmagas = 50\mbox{ g/cm}^2$ at $\rp=10\au$.
Here we consider the observations in range of millimeter and sub-millimeter by e.g., Atacama Large Millimeter/submillimeter Array (ALMA).
In this range of wave length, the opacity of dust grains with $\dsize\sim 1 \mbox{ cm}$ is much 0.1 times smaller than that of the millimeter-sized dust grains \citep[e.g.,][]{Miyake_Nakagawa1993}.
Hence, almost whole region of the dust ring is observable as a bright ring.
Note that if assuming micron-sized grains, we would underestimate the mass of the dust grains.

Until now, the broad dust rings which the width is comparable to its distance from the central star have been observed in several protoplanetary disks \citep[e.g.,][]{Perez2014,van_der_Plas_etal2017}.
Such a broad ring could be induced by the planet when the dust feedback is considered.
As discussed above, once the broad ring is formed, multitude of planets can be formed.
After the formation of the planets, the broad structure could be destroyed by each planet.
In the protoplanetary disk holding multiple rings and gaps such as the disks of HL~Tau, TW~Hydra, MWC~758 \citep{ALMA_HLTau2015,Nomura_etal2016,Dong2018}, hence, it would be difficult to identify the broad structure of the dust ring, even if it was formed before.

\subsection{Validity of the model} \label{subsection:validty_model}
In Section~\ref{subsec:ring_jupiter}, we have shown that the diffusion of the dust grains plays the critical role on the formation of the broad dust ring.
In this paper, we adopt the model of the turbulent diffusion of the dust grains described by Equation~(\ref{eq:j}).
In Equation~(\ref{eq:j}), we assume that the dust grains diffuse only due to gas turbulence.
However, when $\sigmadust \sim \sigmagas$, the turbulence could also be driven by the instability of the dust layer \citep[e.g.,][]{Sekiya1998,Sekiya_Ishitsu2000,Johansen_Henning_klahr2006,Chiang2008,Lee_Chiang_Davis_Barranco2010a,Lee_Chiang_Davis_Barranco2010b}.
This additional turbulence may lead diffusion in addition to that described by Equation~(\ref{eq:j}).
Moreover, Equation~(\ref{eq:j}) is derived from the gradient diffusion hypothesis \citep{Cuzzi_Dobrovolskis_Champney1993,Takeuchi_Lin2002}.
However, the validity of this hypothesis is not confirmed yet when $\sigmadust \gtrsim \sigmagas$.
If the turbulent diffusion of the dust grains is significantly different from that described by Equation~(\ref{eq:j}), the structure of the dust ring may be different from that obtained in this paper.
A more accurate formulation for dust and gas would be required to investigate in future.

We do not consider the formation of the planetesimals in our calculations.
However, within the ring of the dust grains, the planetesimal can be formed via the streaming instability, as discussed in Section~\ref{subsec:ef_formation_planetesimals}.
Only if the efficiency for the formation of the planetesimals is moderate, the wide ring of the dust grains could be formed.
The efficiency for the planetesimal formation after the streaming instability significantly occurs is not fully understood.
In addition, within the dust ring, the value of $\eta$ is much smaller than that of the unperturbed disk.
In such a situation with the small $\eta$, the properties of the streaming instability are also not fully understood.
We also note that the onset of the streaming instability has been investigated in inviscid situations, but it may depend on the viscosity (or turbulent diffusion).
Further investigations are required to understand the properties of the streaming instability in the dusty ring at the outer edge of the planet-induced gap.

We also adopt the simple model of the size evolution of the dust grains provided by \cite{Birnstiel_Klahr_Ercolano2012}.
In this model, we assume that the size distribution of the dust grains is in coagulation--fragmentation equilibrium when the representative size of the dust grains is limited by the fragmentation.
However, at the outer edge of the gap, this assumption may not be valid, because the dust growth timescale is long due to the small dust-to-gas mass ratio.
In this case, it is possible that a part of small grains do not grow up to the representative size and penetrate into the gap.
This non-equilibrium effect may suppress the growth of the dust ring and prevents the depletion of the inner disk of the dust grains.
Moreover, as recently \cite{Dipierro_etal2018} have shown, the feedback from a larger dust grains than that with the representative size may not be negligible, depending on the distribution of the grain size.
The feedback would be more effective if the size of the dust grains is widely distributed, because of the contribution from the larger grains.
To address these issues, the simulations considering multiple size of the dust grains would be required.

\section{Summary} \label{sec:summary}
In this paper, we have investigated the effects of the dust feedback on the structure of the dust ring induced by the planet.
Considering the dust feedback and the turbulent diffusion of the dust grains, we have carried out the two-dimensional gas--dust two fluid hydrodynamic simulations.
Our results are summarized below:
\begin{enumerate}
  \item When the gap is sufficiently deep, the radial flow of the dust grains is almost halted. In this case, we found that the very wide ring of the dust grains can be formed in the outer disk at the outer edge of the gap (Figure~\ref{fig:rings_q1e-3_a4e-3_h0.05_const_dustsize}). 
  The dust feedback from the diffusing dust grains from the ring region modifies the structure of the gas to reduce the gas pressure gradient (and $\eta$) (Figure~\ref{fig:eta_a4e-3_q1e-3}).
  As a result, the dust grains can diffuse from the outer edge of the ring, and the area of the dust ring gradually increases with time. Then the broad ring of the dust grains can be formed when the dust feedback is considered.
  Even when the size evolution of the dust grains is considered, the structure of the ring is quite similar to that in the case the size evolution is not considered if $\alpha\sim 10^{-3}$ and $\ufrag\sim 10\mbox{m/s}$.
In this case, the broad ring can be also formed (Figure~\ref{fig:rings_q1e-3_a4e-3_h0.05_wDustgrowth}).
  \item The increase rate of the area of the dust ring can be estimated by the mass flux of the dust grains from the outside of the disk (Figures~\ref{fig:tevo_ring_a4e-3_q1e-3},\ref{fig:tevo_widths_qvar_a4e-3},\ref{fig:tevo_widths_a1e-3_q1e-3}).
  \item The minimum mass of the planet to form the dust ring is consistent with the pebble-isolation mass provided by \cite{Bitsch_Morbidelli_Johansen_Lega_Lambrechts_Crida2018} in the case of $\alpha \sim 10^{-3}$.
  \item As discussed in Section~\ref{subsec:planet_formation}, if the dust ring is formed by the planet at $\sim 2\au$ in the less massive disk around the Sun-like star, the massive solid core may be able to be formed within the dust ring. This may be connected to the formation of the Hot Jupiters holding the massive solid core like HD~149026b.
  In the disk held by the dwarf star as $\sim 0.1M_{\odot}$, a multitude of Earth-sized planets could be formed within the ring around $0.5\au$, which may explain the formation of the planetary system like TRAPPIST-1.
\end{enumerate} 

\acknowledgements
K.D.K. was also supported by JSPS Core-to-Core Program ``International Network of Planetary Sciences'' and by the Polish National Science Centre MAESTRO grant DEC- 2012/06/A/ST9/00276.
S.O. was supported by JSPS KAKENHI Grant Number JP15H02065 and by MEXT KAKENHI Grant Number JP18H05438.
Y.S. was supported by JSPS KAKENHI Grant Number JP16J09590.
Numerical computations were carried out on the Cray XC30 at the Center for Computational Astrophysics, National Astronomical Observatory of Japan.
\appendix
\section{Gas structures} \label{sec:gas_disk_structures}
\begin{figure*}
	\begin{center}
		\resizebox{0.98\textwidth}{!}{\includegraphics{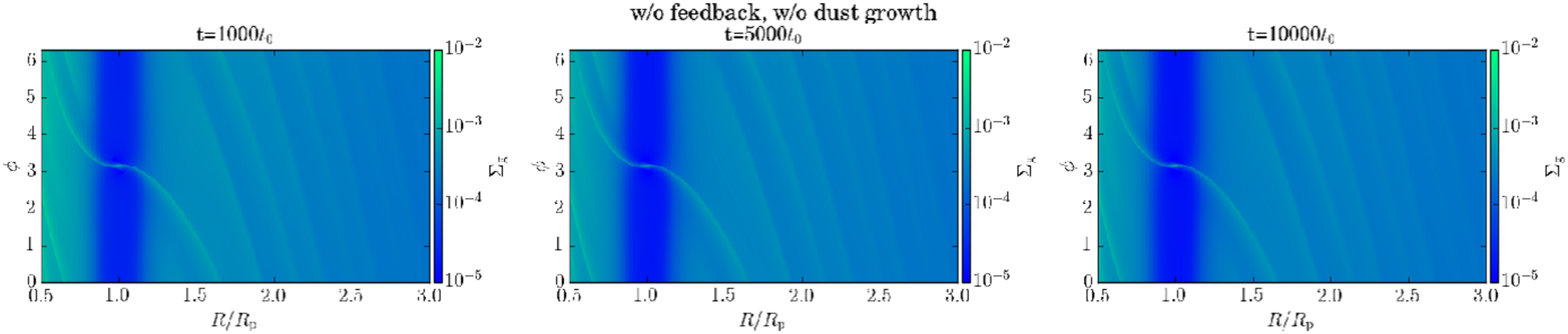}}
		\resizebox{0.98\textwidth}{!}{\includegraphics{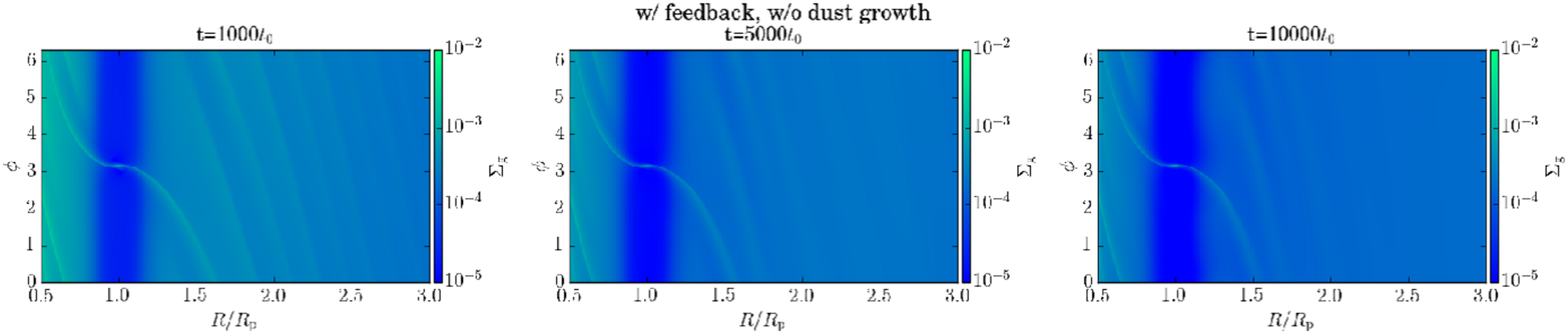}}
		\resizebox{0.98\textwidth}{!}{\includegraphics{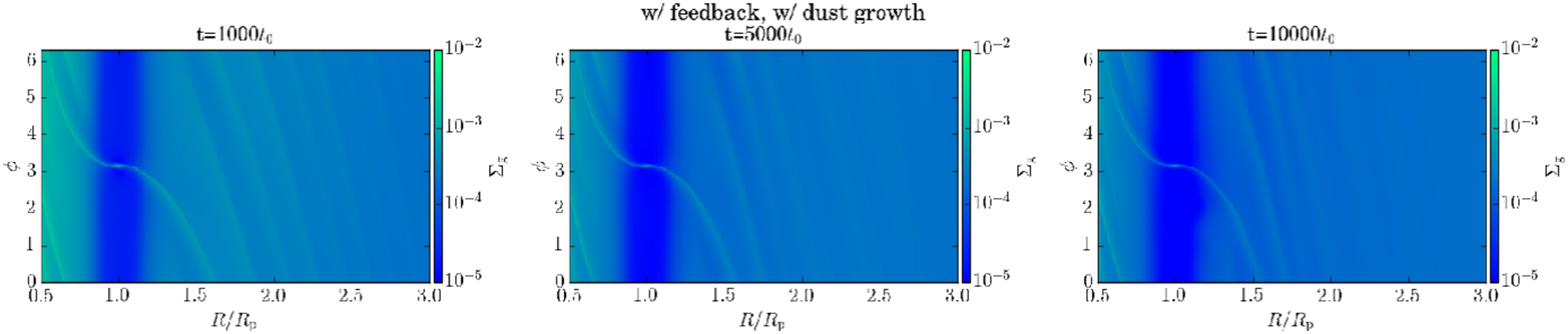}}
		\caption{
		The two-dimensional distributions of the gas surface density in the same as Figure~\ref{fig:rings_q1e-3_a4e-3_h0.05_const_dustsize} and Figure~\ref{fig:rings_q1e-3_a4e-3_h0.05_wDustgrowth}.
		\label{fig:gas_q1e-3_a4e-3_h0.05}
		}
	\end{center}
\end{figure*}
\begin{figure}
	\begin{center}
		\resizebox{0.49\textwidth}{!}{\includegraphics{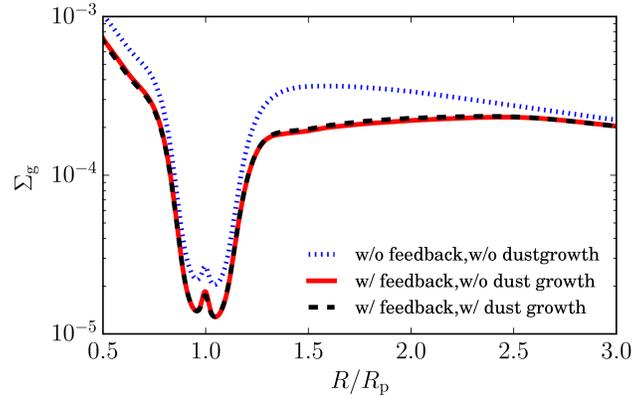}}
		\caption{
		The azimuthally averaged surface density of the gas in the case of Figure~\ref{fig:gas_q1e-3_a4e-3_h0.05}.
		\label{fig:avggasdens_q1e-3_a4e-3_h0.05}
		}
	\end{center}
\end{figure}
When the dust feedback is considered, the structure of the disk gas is modified from that without the dust feedback.
Here we show the gas structure of the disk.
Figure~\ref{fig:gas_q1e-3_a4e-3_h0.05} shows the two-dimensional distributions of the gas surface density when $\mpl/\mstar=10^{-3}$, $\hgp/\rp=0.05$ and $\alpha=4\times 10^{-3}$ (the same case as Figures~\ref{fig:rings_q1e-3_a4e-3_h0.05_const_dustsize} and \ref{fig:rings_q1e-3_a4e-3_h0.05_wDustgrowth}).
No matter whether or not the dust feedback and the dust growth are considered, the distribution of the gas surface density is similar with each other, as different in the structure of the dust grains.
However, the structure of the outer edge of the gap is slightly different.
Figure~\ref{fig:avggasdens_q1e-3_a4e-3_h0.05} shows the azimuthally averaged surface density of the gas at $t=5000t_0$ in the case of Figure~\ref{fig:gas_q1e-3_a4e-3_h0.05}.
As can be seen in the figure, because of the dust feedback, the surface density of the gas at the outer edge of the gap ($1.0 \lesssim R/\rp \lesssim 2.0$) is smaller by a factor of $2$ -- $3$ than that in the case without the dust feedback, whereas the dust growth does not significantly influence the structure of the gas.

\begin{figure}
	\begin{center}
		\resizebox{0.49\textwidth}{!}{\includegraphics{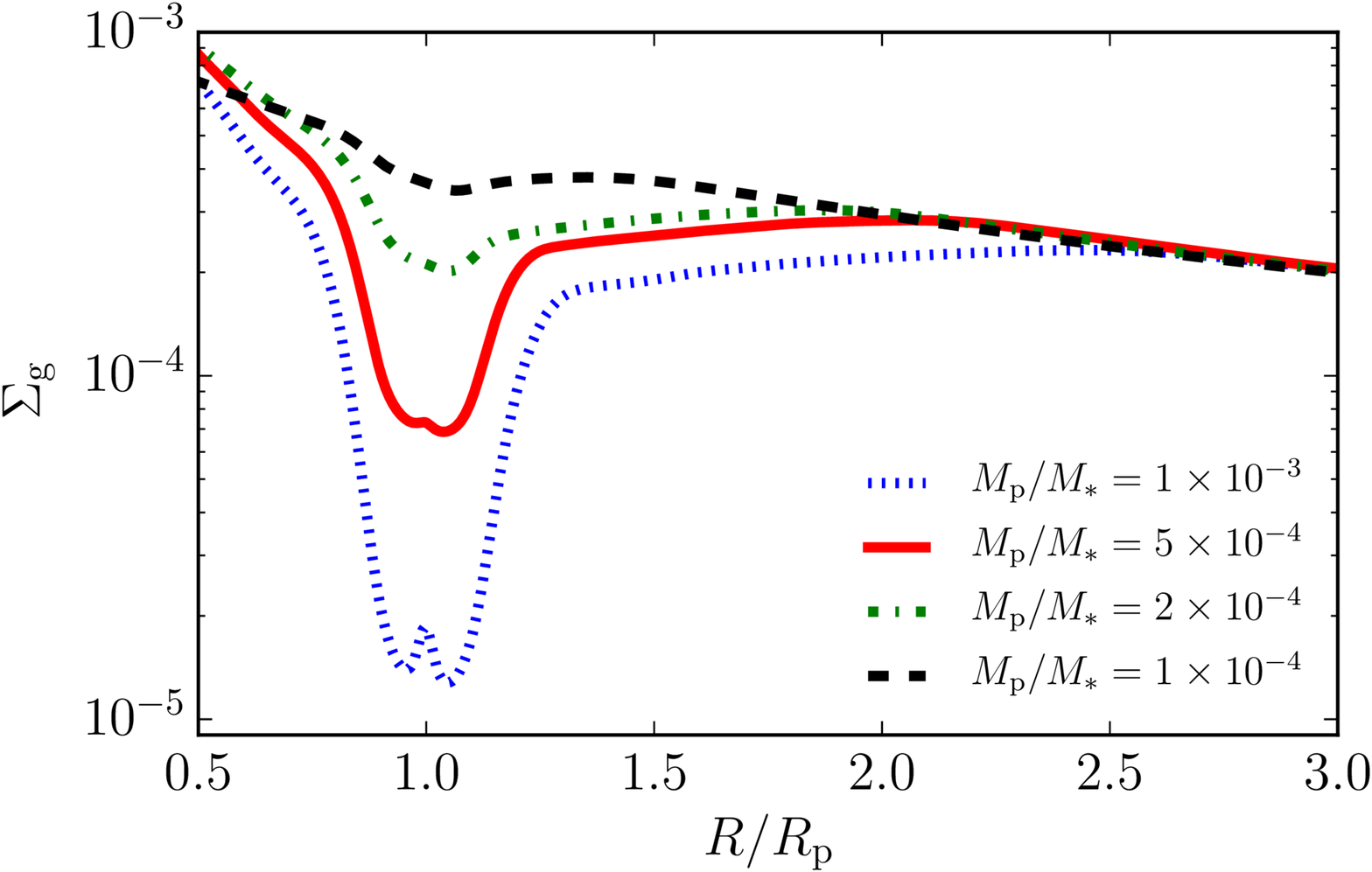}}
		\caption{
		The same as Figure~\ref{fig:avggasdens_q1e-3_a4e-3_h0.05}, but for various planet masses.
		\label{fig:avggasdens_qvar_a4e-3_h0.05}
		}
	\end{center}
\end{figure}
Figure~\ref{fig:avggasdens_qvar_a4e-3_h0.05} shows the azimuthal averaged surface density of the gas at $t=5000t_0$ for the various planet masses when $\alpha=4\times 10^{-3}$ and $\hgp/\rp=0.05$.
As the planet mass increases, the gap becomes deep and the surface density of the gas at the outer edge also decreases.
The surface density of the gas at the outer edge of the gap is larger when the planet mass is smaller or the gap is shallower, though the difference is at the most a factor of two.

\RED{
\section{Resolution convergence} \label{sec:resolution_convergence}
\begin{figure}
	\begin{center}
		\resizebox{0.49\textwidth}{!}{\includegraphics{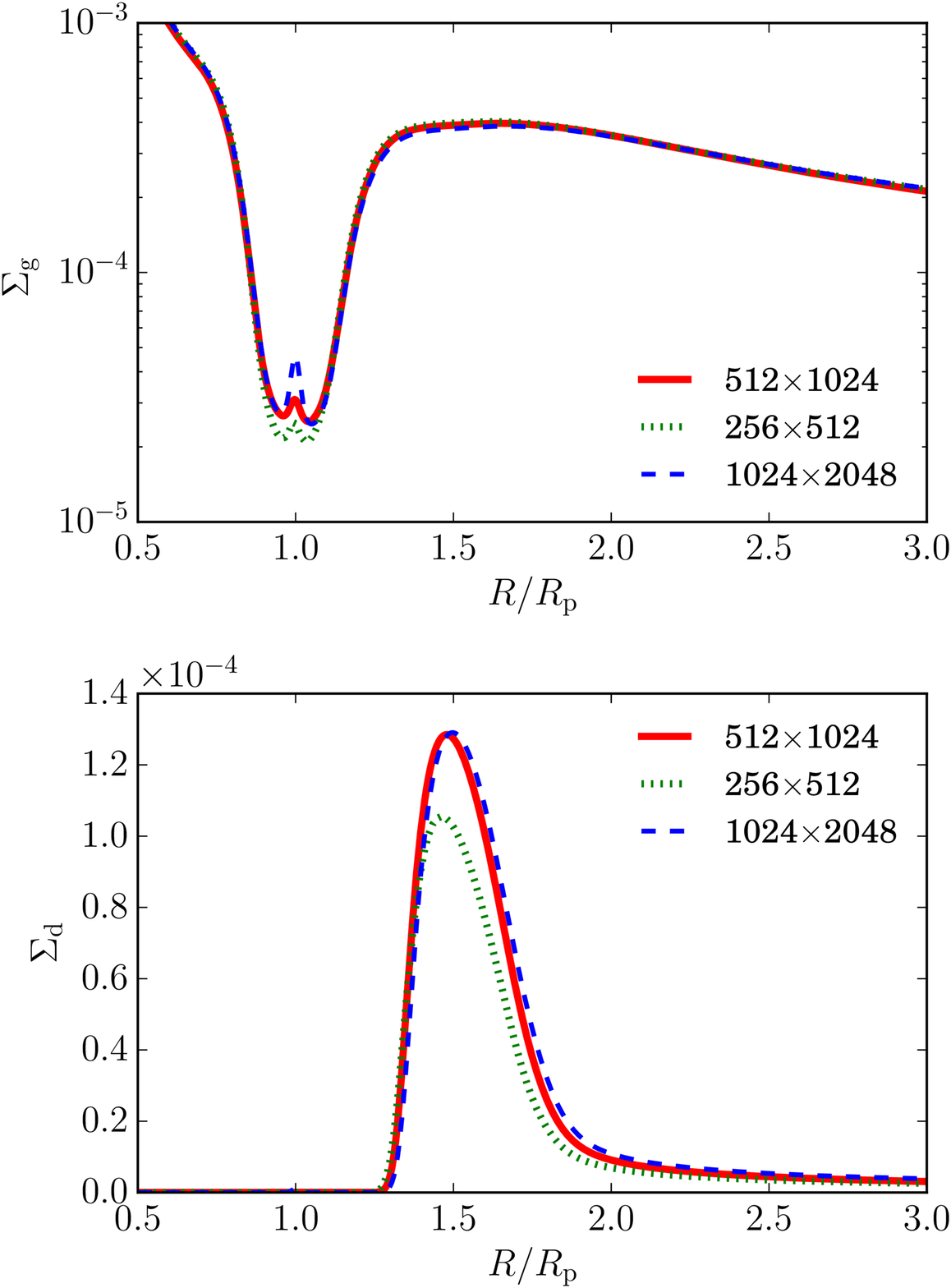}}
		\caption{
		The azimuthally averaged surface density of the gas (upper) and that of the dust grains (lower) at $t=1000t_0$, in the cases of the different grid numbers.
		The numbers of each legend indicates the mesh numbers of radial and azimuthal directions, respectively.
		The parameters are the same as these in Figure~\ref{fig:rings_q1e-3_a4e-3_h0.05_const_dustsize}.
		\label{fig:check_resolution_convergence}
		}
	\end{center}
\end{figure}
In Figure~\ref{fig:check_resolution_convergence}, we show the azimuthally averaged surface densities of the gas and the dust grains, in the same case of that shown in Figure~\ref{fig:rings_q1e-3_a4e-3_h0.05_const_dustsize}, with different resolutions as, ($N_r, N_{\phi}$)= (256,512) (low resolution case), (512,1024) (fiducial case), and (1024,2048) (high resolution case).
As can be seen in the upper panel, the structure of the gas hardly depends on the resolutions.
For the surface density of the dust grains, on the other hand, the surface density of the dust grains within the dust ring around $R\simeq 1.5R_0$ is smaller than these in the fiducial and high resolution cases, because the numerical diffusion is stronger than that in the other cases. 
However, in the fiducial and high resolution cases, the surface densities of the dust grains are quite similar.
Hence, our fiducial resolution is sufficient to resolve the gas structure and the formation of the dust ring.
}

%\bibliographystyle{aasjournal}
%\bibliography{reference}


\end{document}